\documentclass[prc,aps,floatfix,groupedaddress,amsmath,amssymb]{revtex4}
\usepackage{graphicx} 
\usepackage{epsfig,ulem,color} 
\usepackage{dcolumn}
\usepackage{bm}

\def\beq{\begin{equation}}
\def\eeq{\end{equation}}

\newcommand{\rv}{\mathbf{r}}
\newcommand{\rj}{\mathbf{j}}
\newcommand{\kv}{\mathbf{k}}
\newcommand{\qv}{\mathbf{q}}

\newcommand{\kd}{{\rm k}}
\newcommand{\qd}{{\rm q}}

\newcommand{\mbG}{\mathbf{G}}

\newcommand{\mV}{\mathcal{V}}
\newcommand{\mI}{\mathcal{I}}

\newcommand{\vtc}{\tilde{v}_c}

\newcommand{\ra}{\rightarrow}

\newcommand{\be}{\begin{eqnarray}}
\newcommand{\ee}{\end{eqnarray}}
\newcommand{\nn}{\nonumber}
\newcommand{\een}{\nonumber\end{eqnarray}}

\begin{document}

\title{Elementary excitations in homogeneous superfluid neutron star matter: \\ 
role of the proton component}

\author{Marcello Baldo$^1$ and Camille Ducoin$^2$}
\affiliation{
$^1$ Dipartimento di Fisica, Universit\`a di Catania, 
and INFN, Sezione di Catania,
Via Sofia 64, I-95123, Catania, Italy \\
$^2$ Centro de F\'isica Computacional, Departamento de F\'isica, Universidade de Coimbra,
P3004 - 516, Coimbra, Portugal}

\begin{abstract}
The thermal evolution of neuron stars depends on the elementary excitations affecting the stellar matter.
In particular, the low-energy excitations, whose energy is proportional to the transfered momentum,
can play a major role in the emission and propagation of neutrinos.
In this paper, we focus on the density modes associated with the proton component 
in the homogeneous matter of the outer core of neutron stars
(at density between one and three times the nuclear saturation density, where the baryonic constituants
are expected to be neutrons and protons). 
In this region, it is predicted that the protons are superconductor.
We study the respective roles of the proton pairing and Coulomb interaction
in determining the properties of the modes associated with the proton component.
This study is performed in the framework of the Random Phase Approximation,
generalized in order to describe the response of a superfluid system.
The formalism we use ensures that the Generalized Ward's Identities are satisfied.
An important conclusion of this work is the presence of a pseudo-Goldstone mode
associated with the proton superconductor in neutron-star matter.
Indeed, the Goldstone mode, which characterizes a pure superfluid,
is suppressed in usual superconductors due to the long-range Coulomb interaction,
which only allows a plasmon mode.
However, for the proton component of stellar matter, the Coulomb field is screened by the electrons
and a pseudo-Goldstone mode occurs, with a velocity increased by the Coulomb interaction.
\end{abstract}

\maketitle

\section{Introduction}

Homogeneous matter below neutron star (NS) crust is expected to have a proton superfluid component. 
The elementary excitations of the matter affect the whole thermodynamics and long term evolution of the star. 
Since the main components of the matter are neutron, protons, electrons and muons~\cite{shap}, 
the spectral properties of these excitations can have a complex structure.  
Collective modes in asymmetric nuclear matter have been studied previously, 
e.g. in Refs.~\cite{Haensel-NPA301, Matera-PRC49, Greco-PRC67}. 
In the astrophysical context, a study of the collective excitations in normal neutron star matter 
on the basis of the relativistic mean field method has been presented in Ref.~\cite{Providencia-PRC74}. 
The spectral functions of the different components in normal neutron star matter 
have been calculated in Ref.~\cite{paper1,paper2} on the basis of non-relativistic Random Phase Approximation (RPA) 
for the nucleonic components and relativistic RPA for the leptonic components.
Different models for the nuclear effective interaction were considered 
and a detailed comparison was done between some Skyrme forces and a microscopically derived interaction. 
In this work, we will extend this study to the case of superfluid matter.
The elementary excitations in superfluid neutron star matter have been studied by several authors, 
often with controversial results~\cite{Reddy,Kundu,Leinson1,Armen,Leinson2,Vosk}.
One important issue is weither the proton superfluid presents a Goldstone mode at low momentum, since this can have a strong influence e.g. on neutrino emission~\cite{Yako,Reddy,Kundu,Leinson1,Armen,Leinson2,Vosk} or mean free path.
To address this question, we formulate the general theoretical scheme within conserving approximations~\cite{BaymKad,Baym}, which guarantee current conservation and the fulfilment of the related Generalized Ward's Identities (GWI)~\cite{Schriefferb}. 

As it is well known, a neutral superfluid must present a Goldstone mode at low momentum.
For a single charged superfluid, the Coulomb interaction suppresses the mode, which is replaced by the plasmon mode. 
In neutron star matter, the physical situation is complicated by the multi-component structure.
The plasmon mode is in fact mainly an electron excitation; 
furthermore the nuclear interaction couples neutron and proton excitation modes. 
We will focus the study only on the proton and electron components, leaving the complete treatment to future works. 
The reason of this choice is that the role of Coulomb interaction is crucial in this case,
while the coupling with neutrons is expected to be weak~\cite{paper1}. 
What is left out from the treatment is the possible role of the entrainment~\cite{chamel,haensel} 
when both neutrons and protons are simultaneously superfluid. 
However, below the crust, neutron superfluidity rapidly disappears at increasing density
\footnote{We are considering here the outer core of neutron stars. 
At even higher density, the $^3P_2$ neutron pairing channel is expected to play a role.}; 
therefore we will assume that only protons are superfluid.
The aim of the paper is to have a qualitative picture of the excitation spectrum and strength function
associated with the proton superfluid. For simplicity, in the numerical applications, we will consider the effective mass
equal to the bare one, and the nuclear proton-proton interaction will be treated only schematically.

The plan of the paper is the following. 
Section~\ref{form} is devoted to the formalism within the conserving approximation scheme: 
the basic equations are derived and their properties discussed. 
In Section~\ref{pair}, we consider the case of a pure pairing interaction. 
This case is well discussed in the literature, but it is useful to present it before the subsequent analysis.
In a nuclear system, the situation corresponds to strong pairing, 
where the pairing gap can be a substantial fraction of the Fermi energy.
In Section~\ref{pseudo}, we consider the case where both pairing and Coulomb interaction are present, 
and we discuss the interplay of these two interactions in determining the overall structure of the excitation spectrum.
In Section~\ref{strength}, we present the numerical results for the spectral functions in different physical situations.
Conclusions are drawn in Section~\ref{conclu}.

\section{Formalism \label{form}}

Let us first consider a non-superfluid system. We will concentrate on the density fluctuations. 
The correlation function $\Lambda$ is defined by
\be
\label{eq:Lambda}
\Lambda(1,2;3)\, =\, <T(\rho(3)\psi^\dagger(1)\psi(2)> \,,
\ee
where $i \equiv ({\bf r}_i,t_i,\sigma_i,\tau_i)$ stands for coordinate, time, spin and
isospin variables for single particle states; $\rho(3)=\psi^\dagger(3)\psi(3)$ is the spin-isospin density operator.
According to the conserving approximation scheme~\cite{Baym}, $\Lambda$ satisfies the integral equation:
\be
\label{eq:RPA}
\Lambda(12;3)&=& \Lambda_0(12;3) + \Lambda_0(12;\overline{1^{\prime}}\overline{2^{\prime}})
\mV(\overline{1^{\prime}}\overline{2^{\prime}};\overline{4}\overline{5}) \Lambda(\overline{4}\overline{5};3)
\ee
where $\Lambda_0$ is the free correlation function and $\mV$ is the effective particle-hole interaction.
A bar over a symbol $i$ indicates integration and summation over the corresponding set of variables.
$\Lambda_0$ is defined by:
\be 
\label{eq:GG} 
\Lambda_0(12;3)&=&\frac{1}{i}\mbG(13)\mbG(32)\,.
\ee
The quantity $\mbG$ is the single particle Green's function:
\be \mbG(12)\,=\, -i<T\{\psi(1)\psi^\dagger(2)\}> \,,\ee
where $\psi^\dagger , \psi$ are the creation and annihilation operators for the considered particles. 
The effective particle-hole interaction $\mV$ is the key quantity. 
If the approximation has to be conserving, $\mV$ must be expressed by the functional derivative:
\be
\label{eq:Veff}
\mV(12;45)&=&i\frac{\delta\Sigma(12)}{\delta \mbG(45)} 
\ee
where $\Sigma$ is the single particle self-energy, defined according the Dyson's Equation
\be 
\label{eq:Dyson}
\mbG^{-1}(12)&=&\mbG_0^{-1}(12)-U(12)-\Sigma(12) \,.
\ee
$\mbG_0$ is the Green's function for non-interacting particles and $U$ is a possible external single
particle potential, which can be local in space and time. 
In Eq.~(\ref{eq:Veff}), the functional derivative is performed 
considering the Green's function as a functional of the external potential. 
If the derivative is taken at $U=0$, then $\Lambda$ describes the linear response of the system. 
In this framework, if we express the self-energy in terms of the Green's function according to some approximate scheme, the coupled equations (\ref{eq:RPA}) and (\ref{eq:Dyson}) for $\Lambda$ and $\mbG$ (or $\Sigma$) 
define the corresponding conserving approximations. 
It has to be stressed that $\mbG$ and $\Sigma$ must be calculated self-consistently.

The simplest approximations for $\Sigma$ are the Hartree or Hartree-Fock ones, 
where the self-energy is a linear functional of the Green's function. 
In this case, the particle-hole interaction $\mV$ in Eq.~(\ref{eq:Lambda})
is just the bare nucleon-nucleon (NN) interaction.
However, in general, the bare nuclear interaction cannot be used to calculate the linear response, 
because of the hard core typical of the NN interaction. 
One can then introduce effective interactions, derived for instance from Skyrme forces or from a microscopic
procedure, assuming that the interaction can be considered unaffected by the nuclear dynamics at the linear
response level~\cite{paper1}. 
A particular case is the Coulomb interaction, which can be introduced without modifications,
since it contains only a soft core.

The excitations of the proton-electron system in neutron star conditions 
were extensively studied in Refs.~\cite{paper1,paper2}, in the absence of pairing effects. 
It was shown that the electron screening suppresses the proton plasma excitation, 
which is then replaced by a sound-like mode, 
where the proton and electron components move in phase and are both strongly excited. 
On the contrary, the electron plasma excitation is almost unaffected and remains nearly a pure electron mode.

The inclusion of superfluidity can be formally achieved by including in the above scheme 
a further discrete variable $\alpha$, labeling the destruction ($\alpha = 1$) and creation ($\alpha = -1$) operators. 
The collective index becomes $i \equiv ({\bf r}_i,t_i,\sigma_i,\tau_i,\alpha_i) \equiv (x_i,\alpha_i)$. 
Some care must be used in generalizing several relationships from the normal to the superfluid case. 
In particular, the equation defining the inverse Green's function $\mbG(1,2)^{-1}$ must be written:
\beq 
\label{eq:inv} 
\int_{{\bf r}_3 \sigma_3 \alpha_3} \mbG^{-1}(1, x_3\, \alpha_3) \mbG(x_3\, -\alpha_3, 2) \,= \, \delta(1 - 2)
\eeq
The effective interaction now includes both a particle-hole component $\hat V_{ph}$ 
and a particle-particle pairing component $\hat U_{\rm pair}$. 
As a result, the single particle self-energy $\Sigma(1,2)$ contains a normal part, $\Sigma^{n}$, 
and an anomalous part, $\Sigma^{a}$, 
corresponding respectively to $\alpha_{1}\, \neq\, \alpha_{2}$ and $\alpha_{1}\, =\,\alpha_{2}$. 
In the Hartree-Fock (mean field) approximation, they can be written:
\be  \Sigma^{n}(1,2)\, &=&\, - i (1 - \delta_{\alpha_{1}\alpha_{2}})\alpha_{1} \delta(t_{1}-t_{2})\delta({\bf
r}_{1}-{\bf r}_{2}) \delta_{\sigma_{1}\sigma_{2}} \int_{{\bf r}_1' \sigma_1' t_1' } v_{ph}({\bf r}_1-{\bf r}_1')
\mbG(x_1'^+\,\, -1, x_1'\,\, 1)\delta(t_1'-t_1)  \label{eq:self1} \\
\ \ \ &\ & \ \ \ \nn\\
\Sigma^{a}(1,2)\, &=&\, {1\over 2} i \delta(t_1-t_2)\delta_{\alpha_{1}\alpha_{2}} \int_{{\bf r}_1' {\bf r}_2'
\sigma_{1}' \sigma_{2}' t_1'} < x_1 x_2 | \hat U_{\rm pair} | x_1' x_2' >_A \mbG({\bf r}_1' t_1'^+ \sigma_1' \alpha_1 , {\bf
r}_2' t_1' \sigma_2' \alpha_2)\delta(t_1'-t_1)  \label{eq:self2}    
\ee
The superscript $+$ indicates that the time variable has been shifted 
by a positive infinitesimal amount to ensure the correct time ordering. 
The index $A$ indicates the anti-symmetrization of the interaction matrix element.
Having in mind the Coulomb potential case, we have assumed that the particle-hole component $v_{ph}$
is a density-density local interaction, and we have neglected the exchange term;
the generalization to a non-local interaction is straightforward.
The interaction $\mV$ to be used in the RPA equation can be obtained by functional differentiation, 
according to Eq.~(\ref{eq:Veff}).
If, in performing the functional derivative of $\Sigma$,
we assume that $v_{ph}$ and $U_{\rm pair}$ do not depend on $\mbG$,
the approximation is still conserving and the procedure is straightforward. 
More details are given in Appendix~\ref{ap:conserv}.
In the following, we will consider that pairing occurs in the $^1 S_0$ channel, 
so that the corresponding effective interaction will be coupled to zero total spin.

Let us now use the momentum representation, obtained by a usual Fourier transform.
For simplicity, we assume the pairing interaction to have a BCS structure, with a constant pairing
gap $\Delta$ and a cutoff (which corresponds to a contact interaction in coordinate representation).
Then the generalized single particle Green's functions in momentum space are~\cite{Nozieres}:
\be \mbG^{12}(q)&=& \alpha_1\left[\delta_{\alpha_1,-\alpha_2}\delta_{\sigma_1,\sigma_2}G(q)
+\sigma_1\delta_{\alpha_1,\alpha_2}\delta_{\sigma_1,-\sigma_2}F(q)\right] \een
where $q$ is the energy-momentum four-vector $(\qv,\omega)$, and:
\be
G(q)&=&\frac{v_{\qd}^2}{E_{\qd}+\omega-i\eta}-\frac{u_{\qd}^2}{E_{\qd}-\omega-i\eta}\nn\\
F(q)&=&u_{\qd}v_{\qd}\left[\frac{1}{E_{\qd}-\omega-i\eta}+\frac{1}{E_{\qd}+\omega-i\eta}\right]\nn\\
u_{\qd}^2&=&\frac{1}{2}\left(1+\frac{\epsilon_{\qd}-\mu}{E_{\qd}}\right)\nn\\
v_{\qd}^2&=&\frac{1}{2}\left(1-\frac{\epsilon_{\qd}-\mu}{E_{\qd}}\right)\nn\\
\epsilon_{\qd}&=&\frac{\hbar^2{\qd}^2}{2m}\nn\\
E_{\qd}&=&\sqrt{(\epsilon_{\qd}-\mu)^2+\Delta^2} \een
where $\qd=|\qv|$ and $m$ is the particle mass.
The equations become algebraic equations directly for the generalized polarization function $\Pi(\qv,\omega)$:
\beq 
\label{eq:FT}
\int_{r r'} \exp[i(qr - q'r')] \Lambda({ r},{r};{r'})\, =\, \delta(q - q') \mathbf{\Pi}(\qv,\omega)
\eeq
where we have used the four-vector notation $qr\, =\, \qv\cdot{\bf r} - \omega t$.
The spin and alpha indices are implicit.

Following the procedure outlined above, we obtain a system of four coupled equations,
of which only three are independent (one being a linear combination of the others).
The generalized RPA equations for the proton system can then be written as:
\be 
\left(\begin{array}{ccc}
1-X^{pp}_{-}U_{\rm pair} &  - X_{GGC}U_{\rm pair} & -2X_{GF}^{+}v_c \\[0.25cm]
-X_{GGC}U_{\rm pair} & 1- X^{pp}_{+}U_{\rm pair} & -2X_{GF}^{-}v_c \\[0.25cm]
X_{GF}^{+}U_{\rm pair} & X_{GF}^{-}U_{\rm pair} & 1-2X^{ph}_{-}v_c \\[0.25cm]
\end{array}\right)
\left(\begin{array}{c}
\Pi^{(-)}_S \\[0.25cm]
\Pi^{(+)}_S \\[0.25cm]
\Pi^{(ph)}_S\\[0.25cm]
\end{array}\right)
&=& \left(\begin{array}{c}
\Pi^{(-)}_{0,S} \\[0.25cm]
\Pi^{(+)}_{0,S} \\[0.25cm]
\Pi^{(ph)}_{0,S} \\[0.25cm]
\end{array}\right)
\label{eq:RPA1} 
\ee
where the $X$ quantities correspond to different components of the Lindhard function,
when generalized to the superfluid case: they will be presented in the following.
$U_{\rm pair}$, the strength of the pairing interaction, is defined to be positive, 
and it is linked to the pairing gap $\Delta$ by the gap equation:
\be
\Delta &=& U_{\rm pair} \int{\frac{d^3 \kv}{(2\pi)^3}\frac{\Delta}{2E_{\kd}}} 
= U_{\rm pair} \int{\frac{d^3 \kv}{(2\pi)^3} u_{\kd}v_{\kd}}
\ee

To express the generalized Lindhard functions, we have introduced the notation:
\be
\label{eq:Xpp}
X_{\pm}^{pp}&=&\frac{1}{2}\left[X_{GG}^{pp}(q)+X_{GG}^{pp}(-q)\right] \pm X_{FF}(q) \\
\label{eq:Xph}
X_{\pm}^{ph}&=&X_{GG}^{ph}(q) \pm X_{FF}(q) \\
\label{eq:XGF}
X_{GF}^{\pm}&=&X_{GF}(q)\pm X_{GF}(-q) \\
\label{eq:XGGC}
X_{GGC}&=&\frac{1}{2}\left[X_{GG}^{pp}(-q)-X_{GG}^{pp}(q)\right]  
\ee
where the different terms are the following four-dimensional integrals:
\be 
\label{eq:Xinit}
X_{GG}^{ph}(q)&=&\frac{1}{i}\int\frac{dk}{(2\pi)^4}G(k)G(k+q)\;\;;\;\;
X_{GG}^{ph}(-q)=X_{GG}^{ph}(q) \\
X_{GG}^{pp}(q)&=&\frac{1}{i}\int\frac{dk}{(2\pi)^4}G(k)G(-k+q) \\
X_{GG}^{pp}(-q)&=&\frac{1}{i}\int\frac{dk}{(2\pi)^4}G(k)G(-k-q) \\
X_{GF}(q)&=&\frac{1}{i}\int\frac{dk}{(2\pi)^4}G(k)F(k+q) \\
X_{GF}(-q)&=&\frac{1}{i}\int\frac{dk}{(2\pi)^4}G(k)F(k-q) \\
X_{FF}(q)&=&\frac{1}{i}\int\frac{dk}{(2\pi)^4}F(k)F(k+q)\;\;;\;\; X_{FF}(-q)=X_{FF}(q) 
\label{eq:Xfin}
\ee
The explicit expressions of $X_{\pm}^{pp}$, $X_{\pm}^{ph}$ and $X_{GF}^{-}$ are given in Appendix~\ref{ap:lindhard}.

If the alpha indices are explicitly indicated, the polarization tensor can be written:
\beq \mathbf{\Pi}({\bf q},\omega)\, =\, \Pi_{\alpha_1 \alpha_1' ; \alpha_2 \alpha_2'}({\bf q},\omega) \eeq
and the three components of the polarization tensor appearing in Eq.~(\ref{eq:RPA1}) are defined as
\be
\label{eq:Pi-plus}
\Pi^{(+)} &=&\frac{1}{2}\left(\Pi_{11;\alpha\beta}+\Pi_{-1-1;\alpha\beta}\right)  \\
\label{eq:Pi-minus}
\Pi^{(-)} &=&\frac{1}{2}\left(\Pi_{11;\alpha\beta}-\Pi_{11;\alpha\beta}\right)    \\
\label{eq:Pi-ph}
\Pi^{(ph)}&=& \Pi_{-1 1; \alpha\beta}
\een
where the values of the variables $\alpha$ and $\beta$ are generic 
(they are chosen according to the polarization tensor components that have to be calculated).
The index $S$ in Eq.~(\ref{eq:RPA1}) specifies that scalar (zero total spin) excitations are considered. 
The BCS polarization tensor appears on the right hand side of Eq.~(\ref{eq:RPA1}).
It can be expressed in terms of the same quantities $X$ given by Eqs.~(\ref{eq:Xpp}-\ref{eq:XGF}) 
(see Appendix~\ref{ap:lindhard}).

In the so-called constant level density approximation, i.e. exact particle-hole symmetry, the off-diagonal
matrix elements of Eq. (\ref{eq:RPA1}), which couple the first equation with the other two, 
are of order $\qd^2$, and they contribute to the determinant only through terms of order $\qd^4$. 
As a result, to order $\qd^2$, 
the first equation is decoupled from the other two, and actually can be neglected.

Finally one must take into account the Coulomb coupling between protons and electrons. 
Including proton and electron components, the RPA polarization tensor forms a system of three coupled equations:
\be 
\left(\begin{array}{ccc}
1-X^{pp}_{+}U_{\rm pair} & -2X_{GF}^{-}v_c & 2X_{GF}^{-}v_c \\[0.25cm]
X_{GF}^{-}U_{\rm pair} & 1-2X^{ph}_{-}v_c  & 2X^{ph}v_c \\[0.25cm]
      0            &  2X^{e}v_c & 1-2X^{e}v_c \\[0.25cm]
\end{array}\right)
\left(\begin{array}{c}
\Pi^{ (+)}_S \\[0.25cm]
\Pi^{(ph)}_S \\[0.25cm]
\Pi^{(ee)}_S\\[0.25cm]
\end{array}\right)
&=& \left(\begin{array}{c}
\Pi^{(+)}_{0,S} \\[0.25cm]
\Pi^{(ph)}_{0,S} \\[0.25cm]
\Pi^{(ee)}_{0,S} \\[0.25cm]
\end{array}\right)
\label{eq:RPA2} 
\ee
where $X^e$ is the relativistic Lindhard function for electrons~\cite{Jancovici}, calculated in the Vlasov limit, 
and $\Pi^{(ee)}$ is the corresponding part of the polarization tensor involving the electron (density) component. 
These are the equations to be solved to get the proton and electron strength functions 
in the presence of pairing correlations.

To extract the proton density-density polarization tensor $\Pi_{-1 1; -1 1}$, 
it can be of interest to express the electron polarization tensor in terms of the other components, 
and to obtain a reduced set of equations for $\Pi^{ (+)}_S$ and $\Pi^{(ph)}_S$. 
One gets:
\be 
\left(\begin{array}{cc}
1-X^{pp}_{+}U_{\rm pair} & -2X_{GF}^{-}\vtc  \\[0.25cm]
X_{GF}^{-}U_{\rm pair} & 1-2X^{ph}_{-}\vtc   \\[0.25cm]
\end{array}\right)
\left(\begin{array}{c}
\Pi^{ (+)}_S \\[0.25cm]
\Pi^{(ph)}_S \\[0.25cm]
\end{array}\right)
&=& \left(\begin{array}{c}
\Pi^{(+)}_{0,S} \\[0.25cm]
\Pi^{(ph)}_{0,S} \\[0.25cm]
\end{array}\right)
\label{eq:RPA2_screen} 
\ee
where the screened Coulomb interaction is given by
\beq
\vtc \,=\, v_c/( 1 - 2 X^e v_c )
\eeq
This effective interaction is energy dependent. 
If one approximates the polarization function $X^e$ by taking the zero frequency limit at fixed momentum,
$\vtc$ becomes a static screened Coulomb interaction $\vtc^{(s)}$ between protons:
\beq
\label{eq:static}
\vtc^{(s)}(\qd) \,=\, 4\pi e^2/(\qd^2 + \qd_s^2)
\eeq
The screening wave number is given by $\qd_s^2 \,=\, 3(\omega_p^{e}/v_{{\rm F}e})^2$, 
where $\omega_p^{e}$ is the electron plasma frequency and $v_{{\rm F}e}$ is the corresponding Fermi velocity. 
We will refer to this approximation as the "static approximation".

\section{Pairing correlation only \label{pair}}

The simplest situation is when only the pairing interaction is included. 
For the considered case of Eq.~(\ref{eq:RPA2}) this is formally equivalent to put $v_c = 0$, 
and it corresponds e.g. to an uncharged superfluid. 
The corresponding expression for $\Pi^{(ph)}_S$ can be easily obtained 
by solving the remaining two by two algebraic system. 
Replacing the r.h.s. of Eq.~(\ref{eq:RPA2}) by its expression in terms of the quantities $X$,
given in Appendix~\ref{ap:lindhard}, we obtain:
\be 
\label{eq:opair} 
\Pi^{(ph)}_S \,=\, 2\left[ (1-X^{pp}_{+}U_{\rm pair})X^{ph}_{-} -  (X_{GF}^{-})^2U_{\rm pair}
\right]/(1-X^{pp}_{+}U_{\rm pair})
\ee
One can verify, as shown in Appendix~\ref{ap:lindhard},
that the numerator of this expression vanishes at $\qd = 0$ for any non-zero value of $\omega$.
This implies that the response function and the corresponding strength function,
which are even functions of $\qd$,
are proportional to $\qd^2$ for small q for any non-zero value of $\omega$. 
This is a consequence of the conserving approximation we are following, which guarantees the conservation of current: 
the continuity equation implies this property for the density-density response function (see Appendix~\ref{ap:conserv}). 
This result is in agreement with Ref.~\cite{StRe}, 
where the particular expression for the vertex function from Ref.~\cite{Leinson1} was adopted;
however this expression was valid only for $\omega \ll 2\Delta$. 
In any case this RPA property of current conservation is a well known result 
in the theory of superconductors~\cite{Schriefferb}, where only pairing interaction is present, 
and it is also a consequence of gauge invariance.

\begin{figure}[t]
\begin{center}
\includegraphics[width =0.5\linewidth]{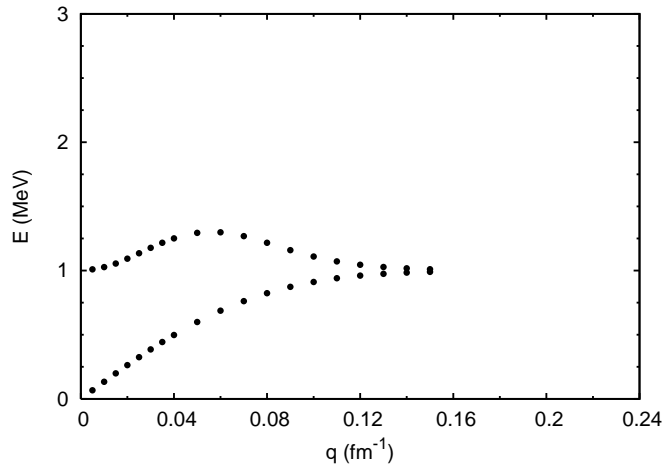}
\end{center}
\caption{Proton excitation spectrum with pairing interaction only.
The proton density corresponds to the case of neutron-star matter at 
nuclear saturation density, with a proton fraction fixed by $\beta$ equilibrium,
which means a proton Fermi momentum $\kd_{{\rm F}p}$ = 0.56 fm $^{-1}$.
The chosen pairing gap is $\Delta \, =\, 0.5\,$ MeV.
}
\label{fig:pair}
\end{figure}

According to a general theorem on symmetry breaking, 
the strength function for $\omega < 2\Delta$ is characterized by the presence of a "Goldstone mode", 
i.e. a phonon-like excitation with an energy proportional to $\qd$ for small enough $\qd$. 
It is well known~\cite{Schriefferb} that in this limit, and in the weak coupling approximation, 
the velocity of the Goldstone mode is $v_{\rm F}/\sqrt{3}$, where $v_{\rm F}$ is the Fermi velocity.
We see on Fig.~\ref{fig:pair} that similar features are obtained in our case.
This figure represents the excitation spectrum of a proton superfluid, in the absence of Coulomb interaction,
for a proton density corresponding to neutron star matter at total baryonic density 
equal to the nuclear saturation density $\rho_0=0.16$ fm$^{-3}$.
The proton fraction (determined by $\beta$ equilibrium in stellar matter) is taken from Ref.~\cite{paper1}: 
it is reported for convenience in Table~\ref{tab:fix}. 
Here we assume for illustration a typical value $\Delta = 0.5$ MeV. 
The exact value of $\Delta$ is not known, 
since the microscopic many-body theory of pairing in nuclear matter is not a settled issue~\cite{Hama}.
The energy of the excitation branches are defined as the zeroes of the determinant of the real part of the
matrix on the left hand side of Eq.~(\ref{eq:RPA2}), for each given value of the momentum $\qd$.  

For small $\qd$ one observes a linear behavior of the Goldstone mode energy;
a numerical measurement of the slope shows that the corresponding velocity is indeed very close to $v_{\rm F}/\sqrt{3}$. 
This velocity can also be calculated according the equation:
\be  1-U_{\rm pair} X^{pp}_{+}\, =\, 0 \label{eq:gold}\ee
which corresponds to the vanishing of the determinant of the matrix 
on the left hand side of Eq.~(\ref{eq:RPA2}). 
Using an expansion of the quantity $X^{pp}_{+}$ up to order $\qd^2$, 
and taking the limit $\Delta \ll E_{\rm F}$, 
we recover from Eq.~(\ref{eq:gold}) the standard Goldstone velocity $v_{\rm F}/\sqrt{3}$.
Let us note that the analytical result depends only weakly on the value of $\Delta/E_{\rm F}$:
this explains why, in our strong coupling case, 
we obtain a Goldstone velocity similar to that of the weak coupling limit.
More details on the analytic expression of the Goldstone mode velocity
are given in Appendix~\ref{ap:expansion}.

According to Eq.~(\ref{eq:opair}), 
the strength function has a delta singularity at the energy of the Goldstone mode, 
since it is undamped in the considered limit of pairing interaction only. 
One can see on Fig.~\ref{fig:pair} that the linear trend is not any more valid when the energy is approaching $2\Delta$.  
Actually the mode can exist only below $2\Delta$, 
since otherwise it could decay in two quasi-particle states and it would be completely damped. 
Above 2$\Delta$ one observes another branch, that starts at  2$\Delta$ for $\qd \,=\, $ 0. 
This excitation can be identified with a "pair breaking" mode. 
A low-momentum development of Eq.~(\ref{eq:gold}) also gives an analytic expression for the energy of this mode,
and shows that it is proportional to $\qd^2$.
However, this quadratic dependence disappears for larger values of $\qd$.
Both branches tend to an energy value close to 2$\Delta$:
this could be expected, since the quantity $X^{pp}_{+}$ has a logarithmic singularity at $\omega \,=\, $2$\Delta$ 
for all reasonable values of the momentum $\qd$, typically for values smaller than the Fermi momentum.
Thus, the two branches nearly touch at a given momentum, where they both disappear. 

\begin{table}[t]
\begin{center}
\begin{tabular}{ccccccc}
\hline
$\rho$ & Y$_p$ & k$_{{\rm F}n}$ & k$_{{\rm F}n}$ & k$_{{\rm F}p}$ & k$_{{\rm F}p}$ & $\omega_{0e}$ \\[0.1cm]
\;[fm$^{-3}$]\; & \;[\%]\; & \;[MeV]\; & [fm $^{-1}$] & \;[MeV]\; & [fm $^{-1}$] & \;[MeV]\; \\[0.1cm]
\hline
0.16 & 3.7 & 327.2 & 1.66 & 110.5 & 0.56 & 6.15 \\
0.32 & 8.6 & 405.2 & 2.05 & 184.1 & 0.93 & 10.25 \\
0.48 & 13.0 & 456.2 & 2.31 & 242.3 & 1.23 & 13.48 \\
\hline
\end{tabular}
\end{center}
\caption{ Density dependence of the proton fraction, Fermi momenta and electron plasmon frequencies in
neutron-star matter conditions. 
The values given for the proton fraction $Y_{p}$ are obtained by microscopic calulation.}
\label{tab:fix}
\end{table}%

\section{Death and resurrection of the (pseudo-)Goldstone mode. \label{pseudo}}

It is a classical result \cite{Schriefferpr} that in a charged superconductor the Goldstone mode cannot be present:
the standard theorem on broken symmetry (the gauge invariance in the present case) 
is not any more valid if the particles interact also by a long range interaction like the Coulomb one. 
The mode is then replaced by a plasmon mode, 
whose frequency is only slightly modified with respect to the case of a normal charged fluid. 
The plasmon branch starts from a non zero energy at $\qd = 0$ (the plasmon energy) 
and varies slowly with the momentum.
This is what we obtain if we consider the first two lines of Eq.~(\ref{eq:RPA2}), 
including the Coulomb interaction $v_c$ but ignoring the electron component:
then we deal with a charged superconductor, the proton liquid. 
In this case, the determinant of the matrix on the left hand side of Eq.~(\ref{eq:RPA2})
vanishes at non-zero energy for $\qd = 0$, since the factor in front of $v_c$ actually vanishes, 
as it happens in a normal charged fluid~\cite{paper1}.

\begin{figure}[t]
\begin{center}
\includegraphics[width =0.5\linewidth]{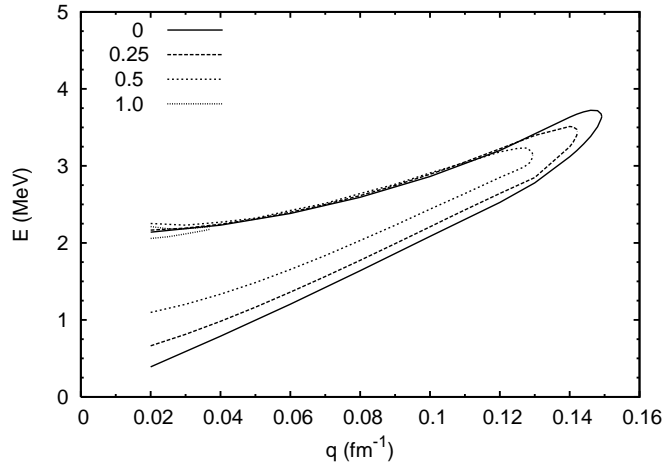}
\end{center}
\caption{
Proton excitation spectrum with pairing and Coulomb interactions, ignoring the electron component.
The proton density corresponds to the case of neutron-star matter at 
nuclear saturation density, with a proton fraction fixed by $\beta$ equilibrium,
which means a proton Fermi momentum $\kd_{{\rm F}p}$ = 0.56 fm $^{-1}$.
The pairing gap takes different values, indicated by the label:
$\Delta$ = 0, 0.25, 0.5 and 1 MeV.
}
\label{fig:plasm}
\end{figure}

The full structure of the excitation spectrum is illustrated in Fig.~\ref{fig:plasm}, 
for the same proton density as in the previous section, and for different possible values of the pairing gap.
When the gap $\Delta$ is much smaller than the plasmon energy,
the structure of the branches resembles the "thumb like" shape 
typical of the charged normal Fermi liquid~\cite{Fetter-Walecka,Jancovici,McOrist,paper1}. 
In this case, the lower branch is over-damped: it does not correspond to an actual excitation. 
The upper branch corresponds to the plasmon mode, 
which is mainly undamped up to a momentum nearby the end of the "thumb", above which no excitation mode exists. 
In the present superfluid case, the lower branch must start above the forbidden energy region $\omega < 2\Delta$.
At increasing value of $\Delta$, the spectrum changes considerably,
although the position of the genuine plasmon excitation remains nearly unchanged:
for a large enough value of the gap, the momentum dependence of the plasmon energy 
deviates more and more from the quadratic form typical of a charged Fermi liquid,
and the whole spectrum shrinks.
When 2$\Delta$ reaches a value close to the plasmon energy $\omega_p$ of the normal proton fluid,
no excitation is possible any more.

\begin{figure}[t]
\begin{center}
\includegraphics[width =1\linewidth]{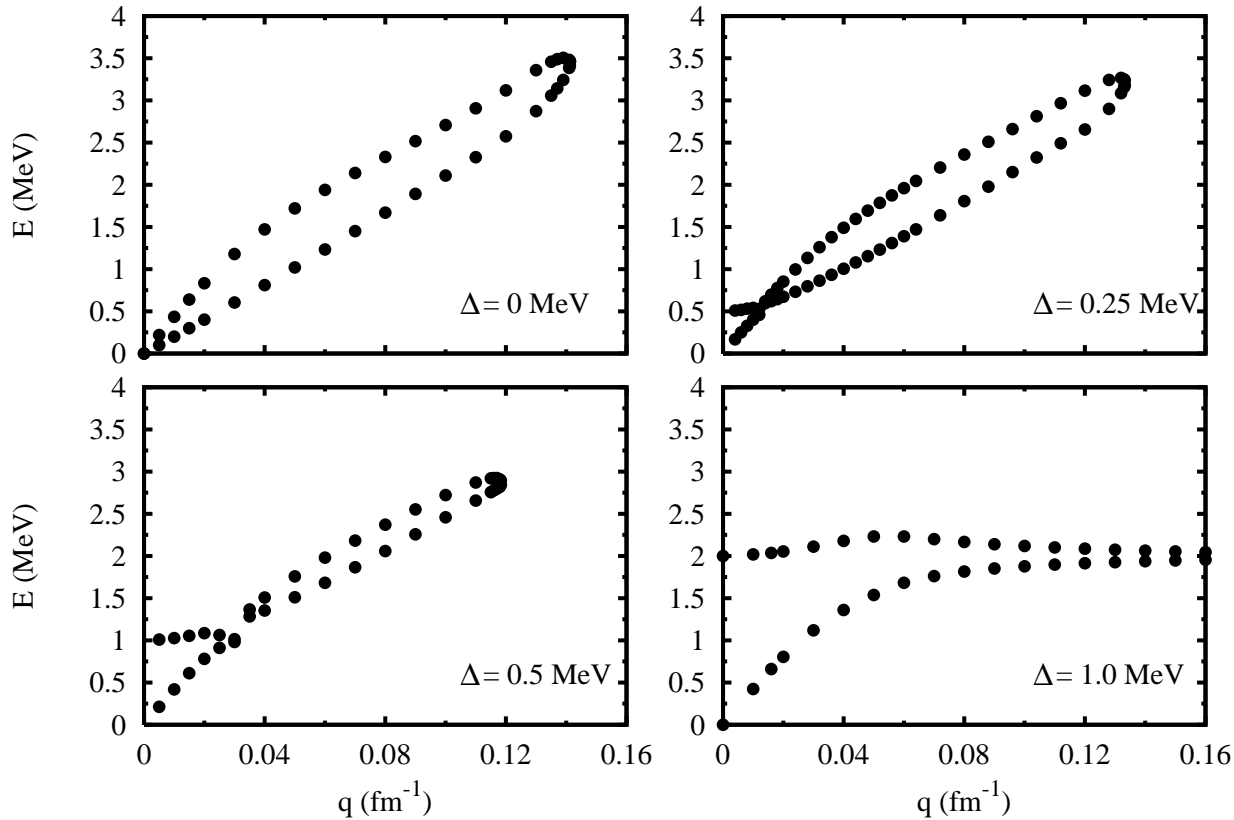}
\end{center}
\caption{
Proton excitation spectrum with pairing and Coulomb interactions, including the electron component.
The proton density corresponds to the case of neutron-star matter at 
nuclear saturation density, with a proton fraction fixed by $\beta$ equilibrium,
which means a proton Fermi momentum $\kd_{{\rm F}p}$ = 0.56 fm $^{-1}$.
The pairing gap takes different values, indicated by the label:
$\Delta$ = 0, 0.25, 0.5 and 1 MeV.
}
\label{fig:branch}
\end{figure}

On the other hand, in a physical situation such as neutron-star matter, 
we have to take into account the presence of the electrons, 
which has a crucial impact on the excitation modes of the medium.
The structure of the spectrum changes radically when the electron component is introduced.
Indeed, electrons are much faster than protons, 
and are able to screen the proton-proton Coulomb interaction at all considered frequencies. 
As a result, the effective proton-proton interaction is of finite range, 
and a sound-like branch appears again. 
This effect was extensively discussed in~\cite{paper1,paper2} in the case of non-superfluid matter.
In the present situation, below $2\Delta$, the sound-like branch associated with the proton component
can be considered as a pseudo-Goldstone mode, which is determined both by pairing and screened Coulomb interaction. 

The structure of the excitation spectrum is illustrated in Fig.~\ref{fig:branch},
for different values of the pairing gap.
In this figure, we report the branches where the strength of the proton component is large.
The plasmon mode, which is mainly electronic, does not appear: it occurs at higher energy. 
The first panel shows the two branches associated with a normal proton fluid, with no pairing. 
The upper branch corresponds to a sound mode, 
which presents some damping, and whose velocity is determined by the screened Coulomb interaction.
The lower branch is over-damped and does not correspond to a true excitation mode.
The two branches join and stop at a cutoff momentum.
When the proton pairing is introduced, the spectrum is still composed of two branches,
but a clear distinction appears between two momentum domains.
For low enough values of $\qd$, one single branch is present below $2\Delta$, 
with an energy approximately proportional to the momentum: this is the pseudo-Goldstone mode.
In this domain, the upper branch is located above $2\Delta$: it corresponds to the pair-breaking mode.
This region of the spectrum is quite similar to the case of pure pairing presented in the previous section,
although the pseudo-Goldstone velocity is different from the pure Goldstone one, as will be discussed in the following. 
For increasing values of $\qd$, these two branches get closer, until a given momentum where they undergo a quasi-crossing.
Afterwards, the upper branch becomes similar to the sound branch of the normal fluid, 
and like for the normal fluid, the lower branch is over-damped and does not correspond to a true excitation.
One can see that the part of the spectrum above 2$\Delta$ tends to shrink as the value of the gap increases. 
For a large enough gap, it actually disappears and only the pseudo-Goldstone and the pair-breaking modes are present,
as illustrated in the last panel of Fig.~\ref{fig:branch}, at $\Delta$ = 1 MeV.

As noticed in the previous section, if only the pairing interaction is considered, 
a Goldstone mode is present below $2\Delta$, with a velocity $v_{\rm G}$ 
that in the weak coupling limit is equal to $v_{{\rm F}p}/ \sqrt{3}$.
The pseudo-Goldstone mode that is obtained in the presence of a screened Coulomb interaction 
has a much higher velocity, about three times larger. 

The excitation spectrum in the region of small values of the momentum $\qd$ can be studied analytically. 
As it is shown in Appendix~\ref{ap:expansion}, 
the expansions of the matrix elements appearing in Eq.~(\ref{eq:RPA2}) 
give an explicit formula for the velocity $v_{\rm PG}$ of the pseudo-Goldstone mode. 
If the small cut-off dependent terms are neglected, 
in the weak coupling limit and static approximation for the electron polarization function 
one gets a quite simple formula:
\beq
\label{eq:vel}
v_{\rm PG}^2 \,=\, v_{\rm G}^2 \left[ 1 \,+\, {N_p\over N_e} 
\left(1 \,-\, {{(\Delta / E_{{\rm F}p})^2}\over 4}\right) \right]
\eeq
where $N_p$ and $N_e$ are the proton and electron level densities.
Since the protons can be considered non-relativistic and the electrons ultra-relativistic, one easily obtains
\beq
\label{eq:den}
{N_p\over N_e} \,=\, {m_p c\over \hbar \kd_{{\rm F}p}}
\eeq
where $m_p$ is the proton mass. 
At the considered density this ratio is about 8, which gives the factor
3 between the pseudo-Goldstone and Goldstone mode velocities.
Equations~(\ref{eq:vel}) and (\ref{eq:den}) show that the ratio between the two velocities decreases 
with increasing density.

\begin{figure}[t]
\begin{center}
\includegraphics[width =0.5\linewidth]{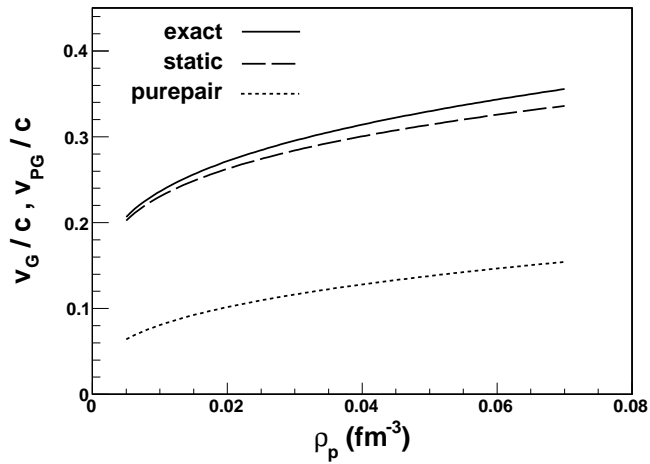}
\end{center}
\caption{
Velocity of the Goldstone mode as a function of proton density.
$v_{\rm G}$ : velocity of the Goldstone mode in the pure pairing case (purepair).
$v_{\rm PG}$ : velocity of the pseudo-Goldstone in the presence of proton pairing and screened Coulomb interaction,
calculated in the static approximation (static) or with an exact treatment (exact).
}
\label{fig:vel_Gm}
\end{figure}

The static approximation for electron polarization function is expected to be quite accurate 
at low momentum and energy.
Figure~(\ref{fig:vel_Gm}) illustrates the relevance of this approximation for the calculation of $v_{\rm PG}$. 
The numerical calculations are in close agreement with the analytical results 
and the agreement is present at all densities. 
Furthermore, we have verified that $v_{\rm PG}/v_{\rm G}$ has a rather weak dependence on the value of the gap.

\section{The strength functions. \label{strength}}

The density strength function $S_v(\qv,\omega)$ is the key quantity for many physical properties
which characterize the behavior and dynamics of neutron star cores, 
like neutrino scattering, neutrino emissions, specific heat and cooling. 
In the present notation, it is given by:
\beq
\label{eq:str}
S_v(\qv,\omega) \, =\, - \Im (\,  \Pi^{(ph)}_S \,  ) \,,
\eeq 
where $\Im$ indicates the imaginary part. 
The function $\Pi^{(ph)}_S $ is also called the vector response function, 
in contrast with the axial vector response function which corresponds to the spin-density strength function.
The latter, which is also a key quantity, can be studied along the same lines;
however, we will concentrate on the vector response function. 
The strength function in the absence of pairing interaction was studied in Ref.~\cite{paper1}.
The pairing correlations have different effects, that we analyze now in detail.

The behavior of the strength function reflects the structure of the excitation spectra
obtained in the previous sections. 
The different peaks that appear correspond to the branches in the spectrum. 
However the position of the peaks in the strength functions can be slightly shifted 
with respect to the corresponding branches. 
This is due to the imaginary part of the response function, i.e. to the damping of the excitations, 
which moves the position of the mode with respect to the branch extracted from the real part only of the RPA matrix.
Furthermore, it can happen that a branch is over-damped and does not correspond to a real mode,
like it is the case for the branch below the normal sound mode in Fig.~\ref{fig:branch}.

\begin{figure}[t]
\begin{center}
\includegraphics[width =0.5\linewidth]{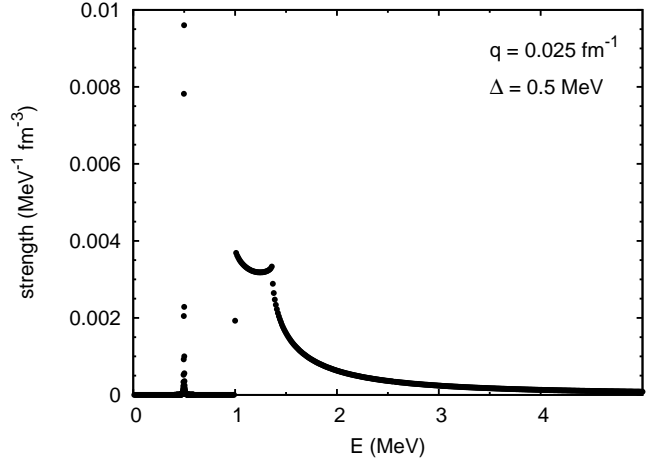}
\end{center}
\caption{
Proton spectral function when only the pairing interaction is included.
The proton Fermi momentum is $\kd_{{\rm F}p}$ = 0.56 fm$^{-1}$.
}
\label{fig:spair}
\end{figure}

For the sake of illustration and analysis, let us first consider, as in Section~\ref{pair}, 
the case of a proton system with only the pairing interaction, and no Coulomb effect.
At not too high momentum, two branches are present in Fig.~\ref{fig:pair}. 
In the corresponding strength function, these two branches give rise to two structures, as shown in Fig.~\ref{fig:spair}, 
where the proton strength function at $\qd$ = 0.04 fm$^{-1}$ is reported. 
The narrow peak below 2$\Delta$ corresponds to the Goldstone mode, which is undamped, 
i.e. it corresponds to a Dirac-delta singularity (numerically one gets just a sharp peak).
Exactly at 2$\Delta$, the spectral function jumps: this corresponds to the opening of the two quasi-particle channel, 
i.e. to the possibility of breaking a Cooper pair.
Above 2$\Delta$, we can observe a structure that corresponds to the pair breanking. 
It is not sharp because it is immersed in the continuum.

\begin{figure}[t]
\begin{center}
\includegraphics[width =0.5\linewidth]{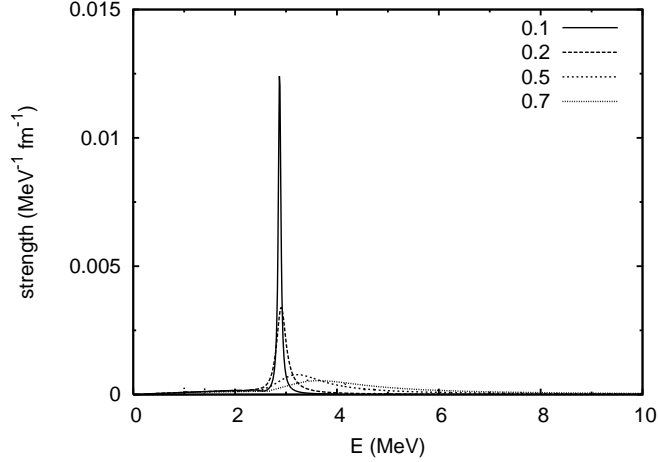}
\end{center}
\caption{
Proton spectral function when pairing and Coulomb interactions are included, without electron screening.
The labels indicate the different values of the pairing gap in MeV. The momentum is $\qd$ = 0.025 fm$^{-1}$.
The proton Fermi momentum is $\kd_{{\rm F}p}$ = 0.56 fm$^{-1}$.
}
\label{fig:scpair}
\end{figure}

We now introduce the Coulomb interaction.
Like in Section~\ref{pseudo},
we first consider the electrons as a static homogeneous background, 
which has no role in the excitation spectrum of the protons.
In this case, the strength function is dominated by the proton plasma excitation above the forbidden region. 
According to Fig.~\ref{fig:plasm}, the proton plasmon mode (upper branch) remains essentially unchanged at low momentum,
but the spectrum shrinks at increasing $\Delta$ and eventually disappears.
The corresponding spectral function is shown in Fig.~\ref{fig:scpair} for $\qd$ = 0.025 fm$^{-3}$, 
and for different values of the gap. 
The plasmon excitation is well pronounced when the pairing gap is small, since then we are close to the normal fluid case. 
As $\Delta$ increases, this mode is more and more damped and shifted.
We can notice that the damping is strong even when the position of the excitation along the branch 
is well below the endpoint of the spectrum (the tip of the "thumb"). 
As regards the lower branch, it is always over-damped, as in the normal system: no strength is associated with it.
We will not develop further the analysis of this case, 
since the role of electrons has to be introduced in order to discuss the realistic physical situation in neutron stars.

\begin{figure}[t]
\begin{center}
\includegraphics[width =1\linewidth]{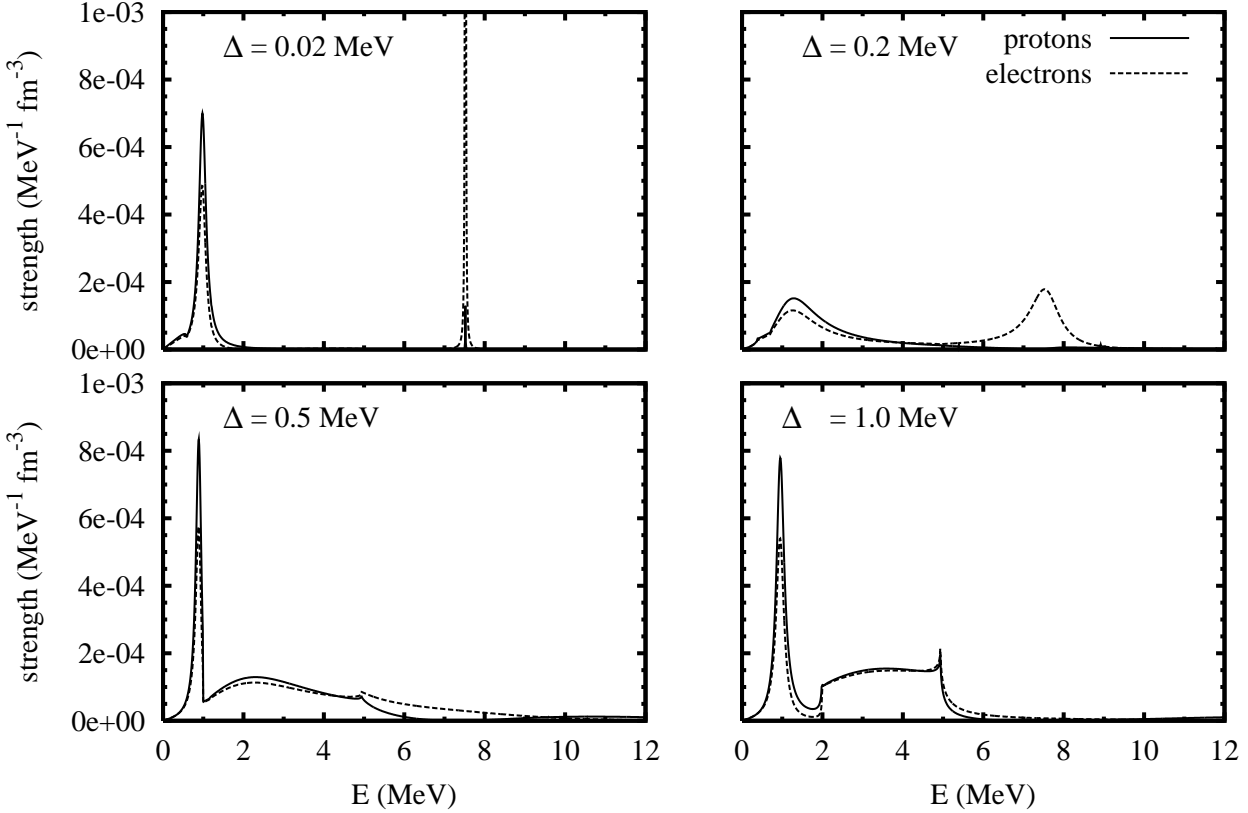}
\end{center}
\caption{
Proton and electron spectral functions, with proton pairing and Coulomb interaction,
for a fixed value of the momentum $\qd$ = 0.025 fm$^{-1}$
and different values of the pairing gap $\Delta$. 
The proton Fermi momentum is $\kd_{{\rm F}p}$ = 0.56 fm$^{-1}$.
}
\label{fig:spec_q0.025}
\end{figure}

Let us now consider the strength function when the electron component is treated dynamically, 
as in a realistic situation.
This case is illustrated in Fig.~\ref{fig:spec_q0.025} for $\qd$ = 0.025 fm$^{-1}$
and for different values of the pairing gap.
When the pairing gap is very small (upper left panel), the system is close to the normal one: 
the spectral function is dominated by the sound mode, well above the so called forbidden region $\omega < 2\Delta $. 
The mode appears as a relatively sharp peak, where the electron strength is comparable with the proton strength. 
This indicates the screening effect of the electrons, 
which are much faster than protons and can easily follow their motion. 
The electron screening is responsible for the transformation of the proton plasmon mode into the sound mode 
when the electron component is treated dynamically. 
The position of the peak can be inferred from the upper branch shown in the first panel of Fig.~\ref{fig:branch}.
The width of the sound mode is determined by the proton-electron coupling, 
since its position is above the proton particle-hole continuum, so that direct Landau damping can not occur. 
As usual, the lower branch is over-damped and no strength is associated with it.

In the upper right panel of Fig.~\ref{fig:spec_q0.025}, the pairing gap $\Delta$ is increased to 0.2 MeV. 
Then, the sound mode falls closer to the forbidden region, but still above it, 
as can be infered from the upper right panel of Fig.~\ref{fig:branch}. 
The mode is strongly damped by the coupling with the two quasi-particle continuum: the Landau damping is now active.

In the lower left panel of Fig.~\ref{fig:spec_q0.025}, $\Delta$ is further increased to 0.5 MeV.
A sharp peak appears again at low energy, but now it is located below $2\Delta$:
this is the pseudo-Goldstone mode. 
In contrast with the Goldstone mode of the pure pairing case, wich corresponds to a Dirac singularity,
the pseudo-Goldstone mode is slightly damped:
this damping is due to the coupling with electrons, since no proton pair breaking can occur in this region. 
Above 2$\Delta$, a broad structure is present, corresponding to the pair-breaking mode.
Although its strength is spread over a large energy interval, 
its total strength turns out comparable with the one of the pseudo-Goldstone mode. 
Further increase of the pairing gap (lower right panel) confirms this trend of the strength function. 

In the last panel, one observes also an abrupt decrease of the strength function at a given energy (about 5 MeV):
this is due to the behavior of the electron free polarization function $X^{e}$ 
(relativistic Lindhard function) involved in Eq.~(\ref{eq:RPA2}). 
Indeed, for $\qd \ll \kd_{{\rm F}p}$, the imaginary part of $X^e$ is proportional to $\qd$,
and contributes to the imaginary part of the particle-hole polarization propagator $\Pi^{ph}$.
In the ultra-relativistic limit \cite{Jancovici}, this contribution suddenly goes to zero at $E = \hbar c \qd$:
this causes the sharp drop of the strength function.
A similar effect, but less pronounced, is observed in the lower left panel. 
This also shows the mutual influence between electron and proton components 
in the excitation modes and the corresponding spectral functions.

The electron strength function, besides following closely the proton strength function at lower energies, 
displays a sharp plasmon mode at very low values of the pairing gap. 
However, as soon as the gap increases, the electron plasmon mode is strongly damped. 
Since the proton Fermi function is smoothed by the pairing interaction, 
we interpret this effect as due to the direct coupling of the electrons with the two proton quasi-particle excitations, whose continuum can extend up to the position of the electron plasmon mode. 
Notice that in this region the position of the electron plasmon mode is well above the electron particle-hole continuum 
and no direct Landau damping is possible. 
The electron plasmon mode actually disappears at values of the gap of few hundreds of keV. 
Such a strong effect is probably overestimated in our calculations, 
where a constant gap has been assumed up to twice the Fermi energy. 
If the pairing gap was assumed to be strongly concentrated around the Fermi momentum, 
the coupling and damping effects would be reduced. 
However, the results clearly display the qualitative behavior of the electron strength function 
as the pairing gap is increased.

\begin{figure}[t]
\begin{center}
\includegraphics[width =1\linewidth]{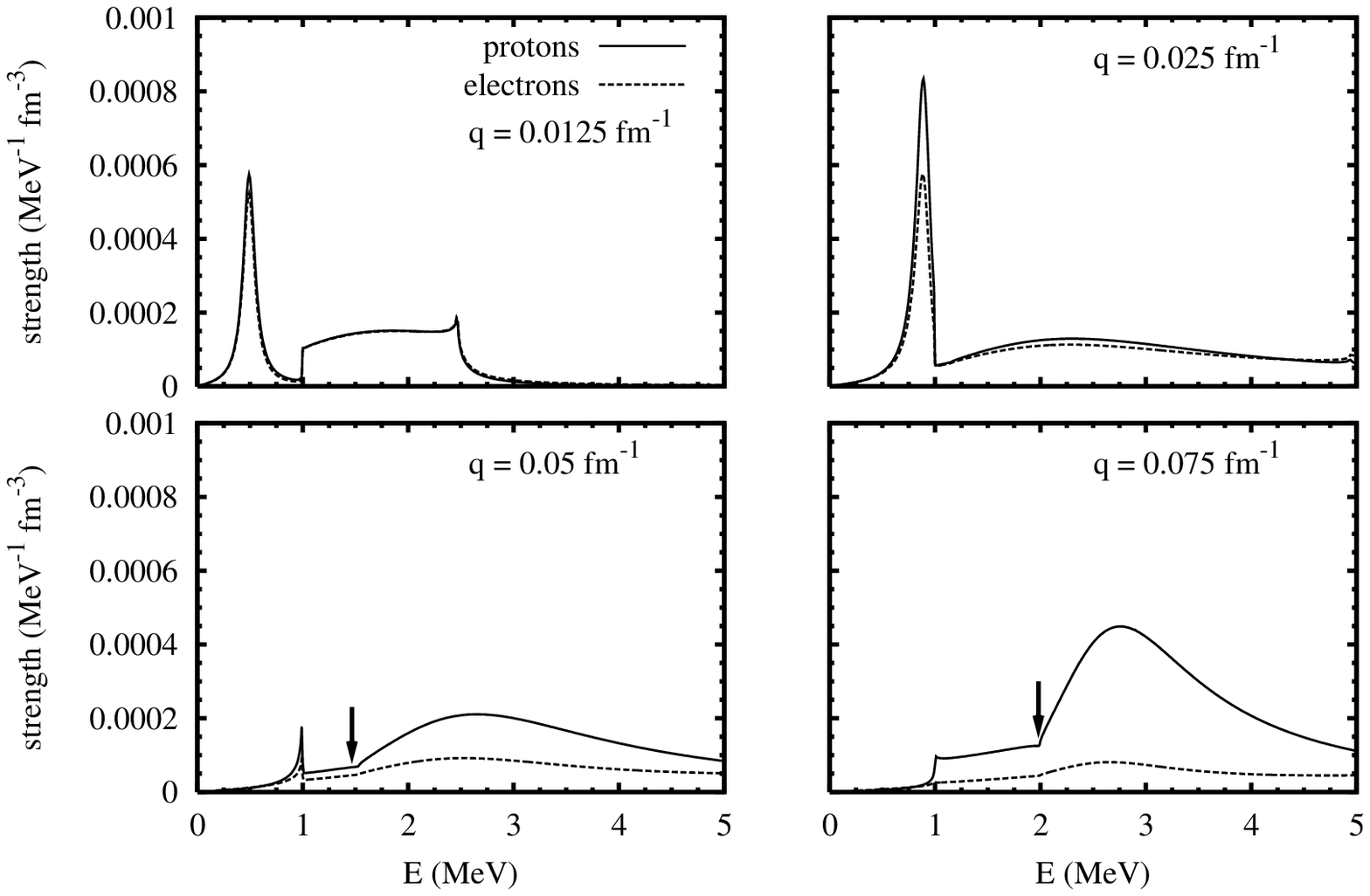}
\end{center}
\caption{
Proton and electron spectral functions, with proton pairing and Coulomb interaction,
for a fixed value of the gap $\Delta = 0.5$ MeV
and different values of the transfered momentum $\qd$.
The proton Fermi momentum is $\kd_{{\rm F}p}$ = 0.56 fm$^{-1}$. 
The arrows point to the kink discussed in the text in relation with Eq.~(\ref{kink}).
} 
\label{fig:spec_d05}
\end{figure}

In Fig.~\ref{fig:spec_d05} is illustrated the development of the proton spectral function, 
which is mainly concentrated at low energy, 
as the momentum increases, for a fixed value of the gap $\Delta = 0.5$ MeV. 
At low momentum (upper left panel), the main mode of excitation is the pseudo-Goldstone mode below 2$\Delta$, 
together with a well defined structure corresponding to the pair-breaking mode, as discussed above.
As the momentum increases (upper right panel), the main peak moves close to 2$\Delta$, but still keeping below,
while the pair-breaking mode is much more spread. 
Notice that still the electron strength function follows closely the proton one. 
Increasing further the momentum (lower left panel), 
the pseudo-Goldstone mode turns into the very damped sound mode above 2$\Delta$, with a residual strength at 2$\Delta$. 
Finally, at higher momentum (lower right panel), the sound mode develops further and becomes slightly more pronounced. 
As the momentum increases, the electron strength is reduced with respect to the proton one, 
since at higher frequency the electron screening becomes less effective.

Another feature of these spectral functions is the presence of two kinks. 
The lowest in energy is trivial: it corresponds to the opening at 2$\Delta$ of the two quasi-particle channels, 
i.e. to the breaking of the Cooper pairs. 
The second one, indicated by an arrow in the lower panels of Fig. \ref{fig:spec_d05}, is less obvious. 
It can be understood from the kink that is present in the two quasi-particle density of states. 
One finds that it occurs at the energy ( $\qd \ll \kd_{{\rm F}p}$ ):
\beq
\label{kink}
E_{\rm kink} \, =\, 2\sqrt{\epsilon_\qd E_{{\rm F}p} \, +\, \Delta^2}
\eeq
With a momentum dependent pairing gap, the kink would be smoothed, but in any case a sudden change of
slope is expected to occur.

\section{Introducing the nuclear interaction}

The effective nuclear particle-hole interaction in the very asymmetric matter 
present in neutron stars is not well known. 
At small momenta, as in all Fermi liquid, it can be characterized by a set of Landau parameters. 
In the language of Landau theory, the density-density interaction $F(\qd)$ 
of two quasi-particles at the Fermi surface is the particle-hole interaction 
to be used in the RPA equations at low momenta. 
It can be expanded in Legendre polynomials as a function of the "Landau angle" 
between the momenta of the two quasi-particles:
\beq
\label{eq:LL}
F(\qd) \equiv F(|\kv_1 - \kv_2|) \, =\, \sum_{L} F_L P_L(cos \theta)
\eeq
where $|\kv_1| = |\kv_2| = \kd_{\rm F}$. 
For asymmetric matter, one has three interaction terms, 
corresponding to neutron-neutron, neutron-proton and proton-proton channels. 
The term $F_0$, corresponding to $L = 0$, is usually dominant. 
In Ref.~\cite{paper1}, it was shown that it can be obtained from the single particle potential 
at the Fermi surface, as obtained from the Brueckner theory. 
If we restrict to the proton-proton channel, the interaction matrix element $V_{\rm nuc}^{pp}$ 
at saturation total baryon density turns out \cite{paper1} to be repulsive and about 311.5 MeV fm$^3$. 
It corresponds to a Landau parameter
\beq
F_0 \,=\, V_{\rm nuc}^{pp} {m \kd_{{\rm F}p} \over \pi^2 \hbar^2} \, \approx\, 0.426 \,.
\eeq

\begin{figure}[t]
\begin{center}
\includegraphics[width =0.5\linewidth]{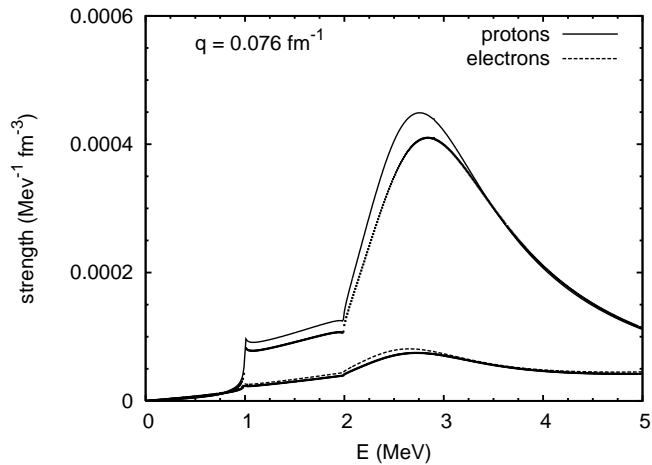}
\end{center}
\caption{
Proton and electron spectral functions, with proton pairing and Coulomb interaction (thin lines);
the thick lines represent these spectral function when the nuclear proton-proton interaction is included.
The pairing gap is $\Delta = 0.5$ MeV and the proton Fermi momentum is $\kd_{{\rm F}p}$ = 0.56 fm$^{-1}$. 
}
\label{fig:nuc}
\end{figure}

The relevance of such an interaction can be estimated by comparing it 
with the screened Coulomb interaction of Eq.~(\ref{eq:static}). 
Even at $\qd$ = 0.1 fm$^{-1}$, the nuclear interaction is less than one fourth of the Coulomb one: 
therefore, only minor corrections are expected to arise from the nuclear proton-proton interaction. 
In fact, one finds that only near the endpoints of the spectrum of Fig.~\ref{fig:branch} 
some corrections to the energies occur, not larger than 10\%. 
As a further illustration, we  show in Fig.~\ref{fig:nuc} the proton and electron spectral functions 
with and without the nuclear proton-proton interaction. 
One can see that the nuclear interaction, being repulsive, 
slightly shifts the proton spectral function to higher energies.
In any case the effect is only marginal.

The effective nuclear interaction can be also calculated from Skyrme forces. 
In a previous paper~\cite{paper1}, 
we have compared the effective forces calculated from the microscopic approach 
with the ones coming from a set of Skyrme forces. 
Despite some qualitative similarities, the effective forces are quantitatively quite different.
The calculation of the effective forces for such highly asymmetric matter 
is a challenge both for the microscopic and the phenomenological approaches.

As for the effective mass, the same microscopic Brueckner calculations 
gives a proton effective mass $m_p^*$ at the Fermi surface of about 0.89 $m$, being $m$ the bare mass. 
It is well known that the inclusion of dispersive dynamical effects enhances slightly the effective mass, 
that is then expected to be close to $m$. 
In conclusion, also the introduction of a realistic proton effective mass can only marginally affect the results
presented in this paper.

\section{Summary and prospects \label{conclu}}

We have presented a study of the elementary excitations in the homogeneous asymmetric nuclear matter
that is present in neutron stars, focusing the analysis on the proton component. 
The protons have been assumed to be superfluid, 
and we have considered the pairing gap $\Delta$ as a parameter taking values up to 1 MeV. 
We have shown how the Goldstone mode that is present in a pure uncharged superfluid below 2$\Delta$
becomes a pseudo-Goldstone mode due to the Coulomb interaction. 
This mode has a velocity about three times larger than the original Goldstone mode. 
We have calculated the spectral functions at different momentum:
at increasing momentum, the pseudo-Goldstone mode merges, through a quasi-crossing mechanism, 
into the sound mode that characterizes the normal system. 
Besides these modes, the pair-breaking mode is clearly apparent in the spectral functions. 
However it has a quite broad structure with a modest strength. 
Finally, we have considered the influence of the nuclear proton-proton interaction. 
It turns out that it has only a weak effects on the overall structure of the spectrum and of the spectral functions.

The main items that still have to be analyzed are the introduction of the neutron components 
and the study of the axial-vector channel, 
with the ultimate aim of calculating the neutrino emission rate and mean free path.
Work is in progress in this direction.


\appendix

\section{Conserving approximations \label{ap:conserv}}

In this appendix, we discuss the conserving approximation for the response function. 
This approach was developed a time long ago by G. Baym \cite{Baym} 
for a normal system, with a central particle-particle interaction. 
The theory can be readily generalized to a more complex interaction and to superfluid systems.
The starting point is the equation of motion for the single-particle Green's function $G$, 
which involves the two-particle Green's function $G_2$. 
Then it is possible to show that if one approximates $G_2$ by a functional of $G$, 
which satisfies a definite functional condition, the average particle density $<n(\rv,t)>$ and current $<\rj(\rv,t)>$, calculated in terms of the single-particle Green's function, satisfy the local conservation law:
\beq
\label{eq:cons}
{\partial \over { \partial t }} <n(\rv,t)> \, +\, \nabla <\rj(\rv,t)> \, =\, 0 
\eeq

This result holds true if
(i) the single particle Green's function satisfies self-consistently the equation of motion with the approximate $G_2$ and (ii) the interaction is local. 
The main reason of this result is the fact that, under the functional condition on $G_2$, 
the contribution of the interaction term to the current vanishes:
the conservation law expressed by Eq.~(\ref{eq:cons}) then follows trivially from the kinetic term contribution. 

For simplicity, we illustrate this point in the case of a pairing interaction
in the mean field (BCS) approximation for $G_2$ in the equation of motion: 
this approximation is known to be conserving. 
Let us take a zero-range pairing interaction, which is the one we have actually used in the present work:
\beq \hat{U}_{\rm pair} \, =\, -{1\over 2}\,\,\, U_{\rm pair}\,\,\, \Sigma_{\sigma,\sigma'}\,\, \int d^3 \rv
\psi^\dagger(\rv,\sigma)\psi^\dagger(\rv,-\sigma) \psi (\rv,-\sigma')\psi (\rv,\sigma')(-1)^{(1-\sigma-\sigma')}
\eeq
where spin-zero pairing has been assumed and $U_{\rm pair}$ ($>0$) is the corresponding strength. 
The relevant component of the Green's function $\mbG(1,1')$ corresponds to $\alpha_1 = 1$ and $\alpha_1' = -1$.
This component will be indicated by $G(x_1,x_1')$. 
The equation of motion for $G$ can be written:
\be 
\label{eq:eqm1} 
i\hbar{\partial\over {\partial t}} G(x_1,x_1') + {\hbar^2\over {2 m}}\nabla_r^2 G(x_1,x_1') &\, =\,&
\delta(t - t')\delta(\rv - \rv')  \\  \nn
&+i& U_{\rm pair}\Sigma_{\sigma'}\,\, <T(\psi^\dagger(\rv -\sigma_1,t)\psi(\rv-\sigma', t)\psi(\rv \sigma', t)
\psi^\dagger(\rv' \sigma_1, t'))> (-1)^{(1-\sigma_1-\sigma')}
\ee
where $x_1 = (\rv,\sigma_1,t)$ and $x_1' = (\rv',\sigma_1,t')$. 
The analogous equation for the derivative on the time $t'$ reads:
\be 
\label{eq:eqm2} 
i\hbar{\partial\over {\partial t'}} G(x_1,x_1') - {\hbar^2\over {2 m}}\nabla_r^2 G(x_1,x_1') &\, =\,&
\delta(t - t')\delta(\rv - \rv')  \\  \nn
&+i& U_{\rm pair} \Sigma_{\sigma}\,\, <T(\psi(\rv \sigma_1, t)\psi^\dagger(\rv' \sigma, t')\psi^\dagger(\rv' -\sigma, t')
\psi(\rv' -\sigma_1, t'))> (-1)^{(1+\sigma_1-\sigma)}
\ee
Summing up the two equations of motion, performing the summation over the spin $\sigma_1$, 
and finally putting $t = t'$ and $\rv = \rv'$, the first two terms on the right hand side 
give the conservation law~(\ref{eq:cons}). 
Let us remind that the density and current can be written in terms of the single particle Green's function:
\be <n(\rv,t)>\, &=&\, \Sigma_{\sigma_1}\, <\psi^\dagger (\rv \sigma_1, t)\psi(\rv \sigma_1, t)>\, =\,
i\Sigma_{\sigma_1} G(x,x^+) \\ \nonumber
<\rj(\rv,t)>\, &=&\, i \Sigma_{\sigma_1}\, {\hbar \over{2m}}(\nabla_\rv - \nabla_{\rv'})
<\psi^\dagger (\rv \sigma_1, t)\psi(\rv' \sigma_1, t)>|_{\rv' = \rv}\, =\, \Sigma_{\sigma_1}
(\nabla_\rv - \nabla_{\rv'})G(x,x^+)|_{\rv' = \rv}
\ee
where $x^+ = (\rv , \sigma_1 , t+\epsilon)$, with $\epsilon$ a positive infinitesimal quantity 
that enables to fix the correct time ordering of the operators in the equal-time limit, 
according to the Green's function definition.

In the mean-field approximation, the interaction contribution 
on the right hand side of Eqs.~(\ref{eq:eqm1}) and (\ref{eq:eqm2}) can be factorized. 
For the first Eq.~(\ref{eq:eqm1}), one has:
\be 
&& <T(\psi^\dagger(\rv, -\sigma_1, t)\psi(\rv, -\sigma', t)\psi(\rv, \sigma', t) \psi^\dagger(\rv', \sigma_1, t'))> 
\nn \\
&\approx& \;
<T(\psi^\dagger(\rv, -\sigma_1, t) \psi^\dagger(\rv', \sigma_1, t'))> <\psi(\rv, -\sigma', t)\psi(\rv, \sigma', t)>
\nn \\
&+& \; <\psi^\dagger(\rv, -\sigma_1, t)\psi(\rv, -\sigma', t)> <T(\psi(\rv, \sigma', t)\psi^\dagger(\rv', \sigma_1, t'))>
\nn \\
&-& \; <\psi^\dagger(\rv, -\sigma_1, t)\psi(\rv, \sigma', t)> <T(\psi(\rv, -\sigma', t)\psi^\dagger(\rv', \sigma_1, t'))>
\ee
A completely analogous factorization applies in the second Eq.~(\ref{eq:eqm2}). 
Taking into account the translational and time reversal symmetries, which imply:
\beq 
<T(\psi^\dagger(\rv -\sigma_1, t)\psi^\dagger(\rv' \sigma_1 t')>\, 
=\, -<T(\psi(\rv -\sigma_1, t)\psi(\rv'\sigma_1, t')>\,,
\eeq
the two factorized $G_2$ are equal. 
Since they are multiplied by an opposite phase in the equations of motion, they cancel out. 
The conservation law for the response function follows 
by using the basic formula for the functional derivative~\cite{Baym}:
\beq
\label{FD}
{\delta G(x,x') \over {\delta U(x_1})} \, =\, -i<T(\psi(x) \psi^\dagger(x') n'(x_1)> 
\eeq
where $U(x_1)$ is an external scalar potential coupled linearly to the density $n(x_1)$, 
and $n'(x_1) = n(x_1) - <n(x_1)>$. 
Indeed, the functional derivation of Eq.~(\ref{eq:inv}) gives
\beq 
\label{eq:der} 
\int_{x_3}
\left[ {{ \delta \mbG^{-1}(1, x_3\, \alpha_3)}\over {\delta U}  } \mbG(x_3\,-\alpha_3, 2) \,+ \,
{ \mbG^{-1}(1, x_3\, \alpha_3)}{{\delta \mbG(x_3\,-\alpha_3, 2)}  \over {\delta U}  }\right]   \,= \, 0
\eeq

By using the chain property of the functional derivative, one gets the integral equation
\be 
\label{eq:GRPA} 
\Lambda(12;3)&=& \Lambda_0(12;3) +\Lambda_0(12;\overline{1^{\prime}}\overline{2^{\prime}})
\mV(\overline{1^{\prime}_{\star}}\overline{2^{\prime}_{\star}};\overline{4}\overline{5})
\Lambda(\overline{4}\overline{5};3)
\ee
where a subscript $\star$ indicates a sign change of the variable $\alpha$,
e.g. $1_{\star} = (\rv_1,\sigma_1,t_1, -\alpha_1) $. 
The function $\Lambda_0$ has the same expression as in Eq.~(\ref{eq:GG}), 
provided the Green's functions are generalized to the superfluid case. 
The correlation function $\Lambda(12;3)$ and the integral equation~(\ref{eq:GRPA}) 
are the generalization to the superfluid case of Eqs.~(\ref{eq:Lambda}) and (\ref{eq:RPA}), respectively. 
If one puts $x_1 = x_2$ and $\alpha_1 = -\alpha_2 = 1$, 
the density-density response function $\Pi^{(ph)}(1,2)$ can be recovered. 
In this way the conserving property verified by $\Lambda$
is directly transferred to the generalized response function, 
provided the Green's function (or self-energy) is calculated self-consistently. 
In the mean field approximation, which generates the RPA approximation for the generalized response function, 
this means that the self-energy appearing in the RPA equation for the response function 
must be calculated self-consistently within the mean field approximation. 
For the pairing part this means that we have to use the BCS Green's functions. 
This property of the RPA equation, when only pairing interaction appears, 
was recognized a long time ago~\cite{Schriefferb}. 

If other interactions than pairing are present, noticeably the particle-hole interaction, 
the argument can be repeated and this RPA property still holds, provided all interactions are local. 
The mean field corrections to the self-energy in the mean field approximation 
amounts to a shift in the chemical potential, if the exchange term is neglected
(this is the Hartree approximation, which is also conserving).
If the interaction is non-local, or if we include the exchange term, 
the local continuity equation~(\ref{eq:cons}) does not hold any-more. 
It can be restored in the effective mass approximation, where the effective mass is momentum and energy independent. 

It has to be noticed that the conservation law for the vertex function $\Lambda$, 
which appears in Eq.~(\ref{eq:GRPA}), is equivalent to the generalized Ward's identities~\cite{Schriefferb}.
The latter imply the local conservation of current,
which for the polarization tensor in momentum-energy
representation can be written
\beq
\omega\,\, \Pi^{(ph)}_{(00)} \, +\, \Sigma_{i=1,3}\ \  q_i\,\, \Pi^{(ph)}_{(i0)} \, =\, 0
\label{eq:current-conserv}
\eeq
where the index value $i = 0$ corresponds to the density component 
and the values $i = 1,2,3$ correspond to the three current components. 
The (0,0) component $\Pi^{(ph)}_{(00)}$ coincides with the polarization function $\Pi^{(ph)}$ of Eq.~(\ref{eq:Pi-ph}). 
For an isotropic system, the second (current) term is proportional to $\qd^2$ for small $\qd$.
Thus, the conservation law implies that, for any value of the energy $\omega$, 
the density-density response function in the limit of small value of $\qd$ vanishes as $\qd^2$.

\section{Self-energy and effective interaction \label{ap:interaction}}

Let us consider the equation of motion for the single-particle Green's function in the superfluid case. 
The nucleon-nucleon hamiltonian can be expressed in general as: 
\be
H&=&H_0+V \nn\\
H_0&=&-\frac{\hbar^2\nabla^2}{2m} \nn\\
V&=&\frac{1}{2}\int d^3x_{1^{\prime}}d^3x_{2^{\prime}}d^3x_{3^{\prime}}d^3x_{4^{\prime}}
<x_{1^{\prime}}x_{2^{\prime}}|v|x_{3^{\prime}}x_{4^{\prime}}>
\psi^{\dagger}(x_{1^{\prime}})\psi^{\dagger}(x_{2^{\prime}})\psi(x_{4^{\prime}})\psi(x_{3^{\prime}}) 
\ee
where the integrations include summation over spin (for simplicity we do not include the isospin). 
All time variables are equal.
The interaction can be an effective one, in which case it is considered static 
and particle number conserving (weak coupling limit).

The equation of motion for the single-particle Green's function 
can be derived generalizing the result of Appendix~\ref{ap:conserv}:
\be 
i\hbar\frac{\partial \mbG(12)}{\partial t_1} =
&-i& \left[ -\alpha_1\frac{\hbar^2\nabla^2_{\rv_1}}{2m}<T\{\psi(1)\psi(2)\}> \right.
\nonumber\\
&&-\frac{\delta_{\alpha_1,1}}{2} \int_{x^{\prime}_i} {<}x_{1}x_{2^{\prime}}|v|x_{3^{\prime}}x_{4^{\prime}}>_A
<T\{\psi_{t_1}^{\dagger}(x_{2^{\prime}})\psi_{t_1}(x_{4^{\prime}})\psi_{t_1}(x_{3^{\prime}})\psi(2)\}>
\nonumber\\
&&+ \left. \frac{\delta_{\alpha_1,{-1}}}{2} \int_{x^{\prime}_i}
<x_{1^{\prime}}x_{2^{\prime}}|v|x_1x_{4^{\prime}}>_{A}
<T\{\psi_{t_1}^{\dagger}(x_{1^{\prime}})\psi_{t_1}^{\dagger}(x_{2^{\prime}})\psi_{t_1}(x_{4^{\prime}})\psi(2)\}>
\right]
\nonumber\\
&+&\hbar\delta(t_1-t_2)\delta(\rv_1-\rv_2)\delta_{\alpha_1,-\alpha_2}\delta_{\sigma_1,\sigma_2} 
\label{eq:motion-B}
\ee
where $t_1' = t_2' = t_3' = t_4' = t_1$, 
the subscript $A$ indicates anti-symmetrization and the integration includes summation over spin. 
Remind that $\psi(2) = \psi(x_2)$ if $\alpha_2 = 1$, and $\psi(2) = \psi^\dagger(x_2)$ if $\alpha_2 = -1$. 
In the Hartree or Hartree-Fock approximation, the two-body Green's functions 
on the right hand side of Eq.~(\ref{eq:motion-B})
can be now factorized in the same fashion as in Appendix~\ref{ap:conserv}. 
This enables to extract the self-energy according to the Dyson's Equation~(\ref{eq:Dyson}):
\be 
\Sigma(1,2)=\frac{i\delta(t_2-t_1)}{2}\int_{x_{i^{\prime}}} &&\left[
2\delta_{\alpha_2,-\alpha_1}<x_1x_{1^{\prime}}|v|x_2x_{2^{\prime}}>_A \left(
\mbG(x_{2^{\prime}},t_1^{+},-1;x_{1^{\prime}},t_1^{-},1)\delta_{\alpha_1,-1} -
\mbG(x_{1^{\prime}},t_1^{+},-1;x_{2^{\prime}},t_1^{-},1)\delta_{\alpha_1,1}\right)
\right.\nonumber\\
&&\left. +\delta_{\alpha_2,\alpha_1}<x_1x_2|v|x_{1^{\prime}}x_{2^{\prime}}>_A
\mbG(x_{1^{\prime}},t_1^{+},\alpha_1;x_{2^{\prime}},t_1^{-},\alpha_1)\right] 
\label{Eq:GSigma} 
\ee
If we now assume that the interaction is the sum of a local particle-hole interaction 
and a local zero-range pairing interaction, 
we recover the expression of Eqs.~(\ref{eq:self1},\ref{eq:self2}) for the self-energy. 
From Eq.~(\ref{Eq:GSigma}), we can derive the effective interaction $\mathcal{V}$, to be used in RPA equations:
\be
\mathcal{V}(1,2;4,5) =\frac{\delta\Sigma(1,2)}{\delta\mbG(4,5)} &=&i\delta(t_2-t_1)\delta_{\alpha_2,-\alpha_1}
\delta_{\alpha_4,-1}\delta_{\alpha_5,1} \left[ <x_1x_5|v|x_2x_4>_A\delta_{\alpha_1,-1} -
<x_1x_4|v|x_2x_5>_A\delta_{\alpha_1,1} \right]
\nonumber\\
&+& \frac{i}{2}\delta(t_2-t_1)\delta_{\alpha_2,\alpha_1} \delta_{\alpha_4,\alpha_1}\delta_{\alpha_5,\alpha_1}
<x_1x_2|v|x_4x_5>_A
\label{Eq:GVeff}
\ee
Again, if we assume that the interaction is the sum of 
a local particle-hole interaction and a local zero-range pairing interaction, 
the expression simplifies and we can easily derive the RPA equations in the energy-momentum representation
that we have used in our calculations: see Eqs.~(\ref{eq:RPA1}) and (\ref{eq:RPA2}).

\section{Generalized Lindhard function and numerical method \label{ap:lindhard}}

The relevant functions $X$ appearing in the kernel of the RPA equations, 
reported in Eqs.~(\ref{eq:Xpp}-\ref{eq:XGF}), after the integrations over energy, can be written:
\be X^{pp}_\pm &=&  \frac{1}{2}\int \frac{d^3\kv}{(2\pi)^3} \left(u_{\kd} u_{|\kv+\qv|} \pm v_{\kd}v_{|\kv+\qv|}\right)^2
\left[ \frac{1}{E_{|\kv+\qv|}+E_{\kd}+\omega-i\eta} +\frac{1}{E_{|\kv+\qv|}+E_{\kd}-\omega-i\eta}
\right]\nn\\
X_{-}^{ph} &=& - \frac{1}{2}\int \frac{d^3\kv}{(2\pi)^3} \left(u_{\kd} v_{|\kv+\qv|} + v_{\kd} u_{|\kv+\qv|}\right)^2
\left[ \frac{1}{E_{|\kv+\qv|}+E_{\kd}-\omega-i\eta}+\frac{1}{E_{|\kv+\qv|}+E_{\kd}+\omega-i\eta}
\right]\nn\\
X_{GF}^-
&=&
\int\frac{d^3\kv}{(2\pi)^3}\; u_{|\kv+\qv|}v_{|\kv+\qv|}
\left[ \frac{1}{E_{|\kv+\qv|}+E_{\kd}-\omega-i\eta} -
\frac{1}{E_{|\kv+\qv|}+E_{\kd}+\omega-i\eta}
\right]\nn\\
&=&
\int\frac{d^3\kv}{(2\pi)^3}\; u_{\kd}v_{\kd}
\left[ \frac{1}{E_{|\kv+\qv|}+E_{\kd}-\omega-i\eta} -
\frac{1}{E_{|\kv+\qv|}+E_{\kd}+\omega-i\eta} \right]\nn \ee
Despite the approximation of constant pairing gap, 
these integrals cannot be calculated analytically and their numerical evaluation requires particular care. 
We follow the method of assigning to the quantity $\eta$ a small value, simulating an infinitesimal, 
and then integrating by an adaptive method the remaining two-dimensional integrals.
We have used the subroutine DT20DQ of the Visual Fortran package.
In this way, both real and imaginary parts can be calculated in few minutes with a simple PC. 
We have found that the results are quite stable if we take $\eta$ values between 10$^{-3}$ and 10$^{-6}$ 
(energies are all calculated in MeV). 
For production calculations we have used the value $\eta$ = 1.5 x 10$^{-5}$.

Finally, for completeness, we give here the expression of the free polarization functions 
appearing on the right hand side of RPA Eq.~(\ref{eq:RPA2}).
In the case we are dealing with, i.e. the density-density response function,
which corresponds to $\alpha = -1$ and $\beta =1$ in Eq.~(\ref{eq:Pi-ph}), we have:
\be
\Pi^{(+)}_{0,S} &=& 2 X_{GF}^- \nn\\
 \  \nn\\
\Pi^{(ph)}_{0,S} &=& 2 X_{-}^{ph} \nn\\
 \  \nn\\
\Pi^{(ee)}_{0,S} &=& 0  \nn \ee
Using these expressions, 
for the pure pairing case (i.e. $v_c \rightarrow 0$) one gets the expression of Eq.~(\ref{eq:opair}) 
for the density-density component $\Pi^{(ph)}_{S}$. 
Then, putting $\qd = 0$ in the previous expressions for the functions $X$, one finds:
\beq
U_{\rm pair}\,\, (X_{GF}^-)^2 \, =\, U_{\rm pair}\,\, \omega^2\, \Delta^2 \left[\int\frac{d^3\kv}{(2\pi)^3}\,\,
\frac{1}{E_{\kd}[(2E_\kd)^2 - \omega^2] } \right]^2 \, =\, \, X_{-}^{ph}\, (1\, -\, U_{\rm pair}\, X^{pp}_+)
\,.
\eeq
Together with Eq.~(\ref{eq:opair}), 
this relation shows that $\Pi^{(ph)}_{S} (\qv,\omega)$ vanishes at $\qd = 0$ for any value of $\omega$:
since it is an even function of $\qd$, 
this implies that it is proportional to $\qd^2$ in the small $\qd$ limit for any $\omega$.
Thus, the condition~(\ref{eq:current-conserv}) resulting from the conservation law is satisfied.
This result can be readily generalized to the case where a finite range particle-hole interaction is present.

\section{Expansion for small momenta \label{ap:expansion}}

A development up to second order in momentum $\qd/\kd_{\rm F}$ is performed.

\subsection{Definitions}

We introduce the variables:
\be
\kappa&=&\kd/\kd_{\rm F}\nn\\
\chi&=&\qd/\kd_{\rm F}\nn\\
\xi&=&\omega/\epsilon_{\kd_{\rm F}}\nn\\
\gamma&=&\Delta/\epsilon_{\kd_{\rm F}}\nn\\
\alpha&=&\frac{\chi^2}{4}-\gamma^2 \ee
In the following, the reduced cutoff momentum will be denoted $\kappa_c$. We also define the functions:
\be e(\kappa,\gamma)
&=&\sqrt{(\kappa^2-1)^2 + \gamma^2} \nn\\
g(\kappa)
&=&\kappa^2-1\nn\\
P(\kappa,\gamma,\xi) &=&
(e-g)^2 \times \left[(2e)^2-\xi^2\right]\nn\\
R(\kappa,\gamma) &=& (e-g)^2 + \gamma^2\nn \ee

It is useful to define the following integral:
\be \mI&=& \int_{g=-1}^{\kappa_c^2-1} \frac{dg}{(2e)^2-\xi^2}\nn \ee
Introducing the variables:
\be
x&=&(e-g)^2-\frac{\xi^2}{2}+\gamma^2\nn\\
x_1&=&x(\kappa_c)\nn\\
x_2&=&x(\kappa=0)=(\sqrt{1+\gamma^2}+1)^2 - \frac{\xi^2}{2} + \gamma^2\nn\\
a&=&\frac{\xi^2}{4}-\gamma^2 \xi^2\nn \ee
the integral $\mI$ becomes:
\be \mI&=&\int_{x_1}^{x_2}\frac{dx}{x^2-a}\nn \ee
so that, defining the variable $b=\sqrt{|a|}$, we have:
\be {\rm if}\; a>0 &:&\;\;\; \mI_{01}=\int_{x_1}^{x_2}{\frac{dx}{x^2-b^2}}
=\frac{1}{2b}\int_{x_1}^{x_2}{dx\left(\frac{1}{x-b}-\frac{1}{x+b}\right)}
=\frac{1}{2b}\left[\ln{\left|\frac{x-b}{x+b}\right|}\right]_{x_1}^{x_2}\\
{\rm if}\; a<0 &:&\;\;\; \mI_{01}=\int_{x_1}^{x_2}{\frac{dx}{x^2+b^2}}
=\frac{1}{b}\left[\arctan\left(\frac{x}{b}\right)\right]_{x_1}^{x_2} \ee

Two approximations will be used to obtain explicit expressions
for the integrals involved in the leading terms of the expansions:\\
(1) $\kappa^2 d\kappa \simeq \kappa d\kappa$, correponding to an integrand peaked around the Fermi momentum ($\kappa=1$);\\
(2) an integral can be neglected if its integrand is an odd function of $\kappa^2-1$ (so that the integral would be exactly zero if its borns were $]-\infty;+\infty[$ instead of $[-1;\kappa_c-1]$).\\
We have checked the accuracy of these approximations.

Note that some of the generalized Lindhard functions presented in this appendix contain a term which is a
diverging integral, namely: $\left[\ln\left|e-g\right|\right]_{\kappa=0}^{\kappa_c}$. However, it cancels with
another term when they are injected in the RPA equation.

\subsection{Expanded expressions of
$X^{pp}_{+}$, $X^{pp}_{-}$, $X^{ph}_{-}$, and $X^{GF}_{-}$}

{Expanded expression of $X^{pp}_{+}$}
\be X^{pp}_{+}&=& -\frac{\kd_{\rm F}^3}{(2\pi)^2\epsilon_{\rm F}} \times \left[
\frac{1}{2}\left[\ln\left|e-g\right|\right]_{\kappa=0}^{\kappa_c} +\frac{\xi^2\mI}{4} +\frac{2\chi^2}{3}
\mI_{\rm ppp} + o(\chi^3) \right] \,. \ee
The integral $\mI_{\rm ppp}$ has the expression:
\be \mI_{\rm ppp} &=& \mI \times \left(\frac{\gamma^2}{4\alpha}\right)
\nn\\
&+& \left[\frac{1}{P}\right]_{\kappa=0}^{\kappa_c}\times \left(3\gamma^2-\frac{5\xi^2}{4}\right)
\nn\\
&+& \left[\frac{x}{P}\right]_{\kappa=0}^{\kappa_c}\times \frac{1}{\alpha^2} \left(-\frac{3\gamma^4}{4} +
\frac{5\gamma^2\xi^2}{16} - \frac{\xi^4}{32}\right)
\nn\\
&+&\left[\frac{1}{P^2}\right]_{\kappa=0}^{\kappa_c}\times
\xi^2\left(-4\gamma^4+3\gamma^2\xi^2-\frac{\xi^4}{2}\right)
\nn\\
&+&\left[\frac{x}{P^2}\right]_{\kappa=0}^{\kappa_c}\times \frac{1}{\alpha} \left(2\gamma^6 -
\frac{9\gamma^4\xi^2}{2} + 2\gamma^2\xi^4 - \frac{\xi^6}{4}\right) \nn \ee
%

{Expanded expression for $X^{pp}_{-}$}
\be X_{-}^{pp} &=&-\frac{\kd_{\rm F}^3}{(2\pi)^2\epsilon_{\rm F}} \times \left[
\frac{1}{2}\left[\ln\left|e-g\right|\right]_{\kappa=0}^{\kappa_c} + \alpha\mI + \mI_{\rm ppm} + o(\chi^3)
\right] \ee
where the integral $\mI_{\rm ppm}$ has the expression:
\be \mI_{\rm ppm} &=& \mI\times \frac{1}{\alpha^2} \left(-\frac{\gamma^6}{\xi^2} + \frac{\gamma^4}{2} -
\frac{\gamma^2\xi^2}{16}\right)
\nn\\
&+&\left[\frac{1}{R}\right]_{\kappa=0}^{\kappa_c}\times \left( -\frac{4\gamma^2}{\xi^2}\right)
\nn\\
&+&\left[\frac{1}{R^2}\right]_{\kappa=0}^{\kappa_c}\times \left(\frac{12\gamma^4}{\xi^2}\right)
\nn\\
&+&\left[\frac{1}{R^3}\right]_{\kappa=0}^{\kappa_c}\times \left( -\frac{8\gamma^6}{\xi^2}
\right)\nn\\
&+&\left[\frac{1}{P}\right]_{\kappa=0}^{\kappa_c}\times \left(-\frac{12\gamma^4}{\xi^2} + 8\gamma^2 -
\frac{5\xi^2}{4}\right)
\nn\\
&+&\left[\frac{x}{P}\right]_{\kappa=0}^{\kappa_c}\times \frac{1}{\alpha^2} \left( \frac{3\gamma^6}{\xi^2} -
2\gamma^4 + \frac{7\gamma^2\xi^2}{16} - \frac{\xi^4}{32}\right)
\nn\\
&+&\left[\frac{1}{P^2}\right]_{\kappa=0}^{\kappa_c}\times \left(16\gamma^6 - 16\gamma^4\xi^2 + 5\gamma^2\xi^4 -
\frac{\xi^6}{2}\right)
\nn\\
&+&\left[\frac{x}{P^2}\right]_{\kappa=0}^{\kappa_c}\times \frac{1}{\alpha} \left(-\frac{8\gamma^8}{\xi^2} +
20\gamma^6 - \frac{25\gamma^4\xi^2}{2} + 3\gamma^2\xi^4 - \frac{\xi^6}{4}\right) \nn \ee
%

{Expanded expression for $X^{ph}_{-}$}
\be X_{-}^{ph} &=&\frac{\kd_{\rm F}^3}{(2\pi)^2\epsilon_{\rm F}} \times \left[ \gamma^2\mI + \frac{2\chi^2}{3}\mI_{\rm phm}
+o(\chi^3) \right] \ee
where the integral $\mI_{\rm phm}$ reads:
\be \mI_{\rm phm} &=& \mI\times \frac{1}{\alpha^2} \left( \frac{3\gamma^6}{\xi^2} - \frac{7\gamma^4}{4} +
\frac{\gamma^2\xi^2}{4}\right)
\nn\\
&+&\left[\frac{1}{R}\right]_{\kappa=0}^{\kappa_c}\times \left( \frac{6\gamma^2}{\xi^2}\right)
\nn\\
&+&\left[\frac{1}{R^2}\right]_{\kappa=0}^{\kappa_c}\times \left(-\frac{12\gamma^4}{\xi^2}\right)
\nn\\
&+&\left[\frac{1}{R^3}\right]_{\kappa=0}^{\kappa_c}\times \left(\frac{8\gamma^6}{\xi^2}\right)
\nn\\
&+&\left[\frac{1}{P}\right]_{\kappa=0}^{\kappa_c}\times \left(\frac{12\gamma^4}{\xi^2}-5\gamma^2\right)
\nn\\
&+&\left[\frac{x}{P}\right]_{\kappa=0}^{\kappa_c}\times \frac{1}{\alpha^2} \left( -\frac{3\gamma^6}{\xi^2} +
\frac{5\gamma^4}{4} - \frac{\gamma^2\xi^2}{8}\right)
\nn\\
&+&\left[\frac{1}{P^2}\right]_{\kappa=0}^{\kappa_c}\times
\left(-16\gamma^6+12\gamma^4\xi^2-2\gamma^2\xi^4\right)
\nn\\
&+&\left[\frac{x}{P^2}\right]_{\kappa=0}^{\kappa_c}\times \frac{1}{\alpha} \left(\frac{8\gamma^8}{\xi^2} -
18\gamma^6 + 8\gamma^4\xi^2 - \gamma^2\xi^4\right) \nn \ee
%

{Expanded expression for $X^{-}_{GF}$}
\be X^{-}_{GF} &=&\frac{\kd_{\rm F}^3}{(2\pi)^2\epsilon_{\rm F}} \times \gamma\xi \left[ \frac{\mI}{2}
+\frac{4\chi^2}{3}\mI_{\rm GFm} + o(\chi^3) \right] \ee
where the integral $\mI_{\rm GFm}$ reads:
\be \mI_{\rm GFm} &=& \mI\times \frac{1}{\alpha^2} \left( \frac{\gamma^4}{4\xi^2} - \frac{3\gamma^2}{16} +
\frac{\xi^2}{32}\right)
\nn\\
&+&\left[\frac{1}{R}\right]_{\kappa=0}^{\kappa_c}\times \left( \frac{2\gamma^2}{\xi^4}\right)
\nn\\
&+&\left[\frac{1}{P}\right]_{\kappa=0}^{\kappa_c}\times \left(\frac{\gamma^2}{\xi^2} - \frac{3}{4}\right)
\nn\\
&+&\left[\frac{x}{P}\right]_{\kappa=0}^{\kappa_c}\times \frac{1}{\alpha^2} \left( -\frac{2\gamma^6}{\xi^4} +
\frac{5\gamma^4}{4\xi^2} - \frac{5\gamma^2}{16} + \frac{\xi^2}{32}\right)
\nn\\
&+&\left[\frac{1}{P^2}\right]_{\kappa=0}^{\kappa_c}\times \left( -4\gamma^4 +3\xi^2\gamma^2 -\frac{\xi^4}{2}
\right)\nn\\
&+&\left[\frac{x}{P^2}\right]_{\kappa=0}^{\kappa_c}\times \frac{1}{\alpha} \left(\frac{2\gamma^6}{\xi^2} -
\frac{9\gamma^4}{2} + 2\gamma^2\xi^2 - \frac{\xi^4}{4}\right) \nn \ee

\subsection{Velocity of the Goldstone mode and pseudo-Goldstone mode}

In the case of a pure pairing interaction, the real part of the determinant 
of the RPA matrix~(\ref{eq:RPA2}) diverges if:
\be
\label{EQ:branche_G}
1-X^{pp}_{+}U_{\rm pair}&=&0
\ee
which reads
\be
\label{EQ:branche_G-bis}
\frac{\xi^2}{4}\mI
+\frac{2\chi^2}{3}\mI_{\rm ppp}
+o(\chi^3)&=&0
\ee
In the limit of vanishing $\xi$, this gives:
\be
\frac{\xi^2}{\chi^2}&=&-\frac{8}{3}\frac{\mI_{\rm ppp}(\xi\ra 0)}{\mI(\xi\ra 0)}
\ee
with
\be
\mI(\xi\ra 0)&=&-\left[\frac{1}{R}\right]\\
\mI_{\rm ppp}(\xi\ra 0)
&=&
-\frac{1}{2}\left[\frac{1}{R}\right]
+3\gamma^2\left[\frac{1}{R^2}\right]
-2\gamma^4\left[\frac{1}{R^3}\right]
\ee
so that:
\be
\frac{\xi^2}{\chi^2}
&=&
-\frac{4}{3}
+8\gamma^2\frac{\left[1/R^2\right]}{\left[1/R\right]}
-\frac{16}{3}\gamma^4\frac{\left[1/R^3\right]}{\left[1/R\right]}
\ee
For $\gamma \ll 1$ we have:
\be
\left[1/R\right] &\ra& -\frac{1}{\gamma^2}\;\;\;;\;\;\;
\left[1/R^2\right]\ra-\frac{1}{\gamma^4}\;\;\;;\;\;\;
\left[1/R^3\right]\ra-\frac{1}{\gamma^6}\;\;\;;\;\;\;
\frac{\xi^2}{\chi^2}\ra\frac{4}{3}
\ee
which leads to the Goldstone velocity:
\be
v_{\rm G}&=&\frac{\omega}{\qd}=\frac{\hbar\kd_{\rm F}}{\sqrt{3}m}
\ee

Let us now consider the pseudo-Goldstone mode,
which is present for the proton superfluid when the Coulomb interaction is taken into account,
but screened by the electron component.
In the static limit for the electron free response function,
we can express the Coulomb interaction between protons as:
\be
\bar{v}_c&=&\frac{v_c}{1+\Pi_{0e}v_c}
\ee

To express the determinant of the RPA matrix, it is useful to introduce the reduced Coulomb interaction $\nu_c$ such that:
\be
v_c&=&\frac{4\pi e^2}{\qd^2}=\frac{\nu_c}{\chi^2} \;\;;\;\; \nu_c=\frac{4\pi e^2}{\kd_{\rm F}^2}
\ee
The corresponding reduced screened Coulomb interaction is:
\be
\frac{\bar{\nu_c}}{\chi^2}&=&\frac{\nu_c}{\chi^2+Q^2}=\frac{\nu_c}{Q^2}(1-\frac{\chi^2}{Q^2}+o(\chi^3))
\ee
where $Q$ is related to the screening length of the electrons.
The Coulomb interaction now enters the expression of the determinant $D$ of the RPA matrix~(\ref{eq:RPA2}) as 
$\frac{\nu_c}{Q^2}(1-\frac{\chi^2}{Q^2}+o(\chi^3))$,
and we obtain:
\be
D&=&
\xi^2\left(\frac{\mI}{4}\right)
+\chi^2\left(
\frac{2}{3}\mI_{\rm ppp}
+\frac{2}{3}\frac{\nu_c}{Q^2}\gamma^2\mI\mI_{\rm ppp}\right)
+o(\chi^2\xi^2)
\,.
\een
In the limit $(\chi,\xi)\ra(0,0)$, the determinant vanishes for
\be
\frac{\xi^2}{\chi^2}&=&
-\frac{8\mI_{\rm ppp}}{3\mI}\left[1+\frac{\nu_c\gamma^2}{Q^2}\mI \right]
\ee
where the term
\be
\frac{\nu_c\gamma^2}{Q^2}\mI 
\ee
gives the Coulomb correction to the velocity of the Goldstone mode.


\end{document}